\documentclass[12pt]{article}
\usepackage{graphicx}
\usepackage{amssymb}
\usepackage{sectsty}
\usepackage{amsfonts}
\usepackage{amsmath}
\usepackage{geometry}
\usepackage{setspace}
\usepackage{portland}
\usepackage{scalefnt}
\usepackage{lineno}
\usepackage{indentfirst}
\usepackage{fancyhdr}
\usepackage{caption}
\usepackage{float}
\usepackage{color}
\usepackage[none,default]{harvard}

\input{tcilatex}
\begin{document}

\begin{center}
\ \bigskip

\begingroup%
%TCIMACRO{\TeXButton{TeX field}{\scalefont{2}}}%
%BeginExpansion
\scalefont{2}%
%EndExpansion
\textbf{A} \textbf{Bayesian Nonparametric Meta-Analysis Model}\footnote{%
This research is supported by grant SES-1156372 from the National Science
Foundation, program in Methodology, Measurement, and Statistics.}\endgroup%
\bigskip

\textbf{\begingroup%
%TCIMACRO{\TeXButton{TeX field}{\scalefont{1.25}}}%
%BeginExpansion
\scalefont{1.25}%
%EndExpansion
George Karabatsos\footnote{%
Corresponding author, Professor George Karabatsos, Department of Educational
Psychology, Program in Measurement, Evaluation Statistics, and Assessments,
College of Education, University of Illinois-Chicago\smallskip , 1040 W.
Harrison St. (MC\ 147), Chicago, IL\ 60607. E-Mail: georgek@uic.edu,
gkarabatsos1@gmail.com.}, Elizabeth Talbott\footnote{%
Associate Professor and Chair, Department of Special Education, College of
Education, University of Illinois-Chicago\smallskip . E-mail:\
etalbott@uic.edu.}, \&\ Stephen G.\ Walker\footnote{%
\noindent Professor of Statistics, The University of Texas at Austin,
Division of Statistics and Scientific Computation. E-mail:\
s.g.walker@math.utexas.edu}\endgroup\bigskip }
\end{center}

\noindent \textbf{In a meta-analysis, it is important to specify a model
that adequately describes the effect-size distribution of the underlying
population of studies. The conventional normal fixed-effect and normal
random-effects models assume a normal effect-size population distribution,
conditionally on parameters and covariates. For estimating the mean overall
effect size, such models may be adequate, but for prediction they surely are
not if the effect size distribution exhibits non-normal behavior. To address
this issue, we propose a Bayesian nonparametric meta-analysis model, which
can describe a wider range of effect-size distributions, including unimodal
symmetric distributions, as well as skewed and more multimodal
distributions. \textbf{We demonstrate our model through the analysis of real
meta-analytic data arising from behavioral-genetic research. }We compare the
predictive performance of the Bayesian nonparametric model against various
conventional and more modern normal fixed-effects and random-effects
models.\bigskip }

\textbf{Short title:}\ \ Bayesian Nonparametric Meta-Analysis\bigskip
.\newpage

\section{\textbf{Introduction\label{Section: Introduction}}}

A research synthesis aims to integrate results from empirical research so as
to produce generalizations (Cooper \& Hedges, 2009). Meta-analysis, also
referred to as the analysis of analyses (Glass, 1976), provides a
quantitative synthesis of statistics that are reported by multiple research
studies. Specifically, each study reports an effect-size statistic, and
provides information about its sampling variance, while the features of the
study may be described by one or more covariates. Typical examples of
effect-size statistics include the unbiased standardized mean difference
between two independent groups (Hedges, 1981), among others
(Konstantopoulos, 2007). Given a sample of effect-size data, the primary aim
of a meta-analysis is to infer the overall effect-size distribution from the
given study population, as well as to infer the heterogeneity of the effect
sizes. Conventional summaries of the overall effect-size distribution
include the mean, which is often referred to as the "overall effect-size,"\
and the variance which describes the heterogeneity of reported effect sizes
in the overall effect-size distribution. Heterogeneity can be further
investigated in a meta-regression analysis, in order to investigate how the
mean effect-size relates to key study-level covariates (e.g., Berkey et al.,
1995; Thompson \&\ Sharp, 1999; Thompson \&\ Higgins, 2002; Higgins \&\
Thompson, 2004). Also, meta-regression analysis can be used to investigate
and test for publication bias in the data, by relating effect-sizes with
their standard errors (precisions; the square-root of the effect-size
sampling variances) (Thompson \&\ Sharp, 1999). This actually provides a
regression analysis for the funnel plot (Egger et al., 1997).

The normal fixed-effects model and the normal random-effects model provide
two traditional and alternative approaches to meta-analysis (Hedges \&\
Vevea, 1998; Konstantopoulos, 2007; Borenstein et al. 2010). Each model is a
weighted linear regression model, which treats study-reported effect-size as
the dependent variable, weighs each reported effect-size by the inverse of
its sampling variance, assumes normally-distributed regression errors, and
represents the overall (mean) effect size by the intercept parameter. In
other words, each of these models assume that the effect-size distribution
is a unimodal and symmetric, normal distribution, conditionally on all model
parameters, and conditionally on any set of chosen values of covariates in
the model. The fixed-effects model is an ordinary (weighted)\ linear
regression model. The normal random-effects model is a two-level model that
extends the fixed-effects model, by allowing for between-study variance in
the effect-sizes, through the addition of random intercept parameters that
are assumed to have a normal distribution over the given study population.
Specifically, effect-sizes at the first level are nested within studies at
the second level. A three-level meta-analysis model can accommodate data
structures where, for example, studies (level 2) are themselves nested
within units, such as school districts (level 3) (for more details, see
Konstantopoulos, 2011). Given a sample set of meta-analytic data, the
parameters of a normal fixed-effects or normal random-effects model can be
estimated via maximum-likelihood. Alternatively, a Bayesian inference
approach can be taken, which involves the specification of a prior
distribution on all parameters of the given meta-analytic model (e.g.,
Higgins et al., 2009). Then, Bayesian inference of the data is based on the
posterior distribution of the parameters, formed by combining the data
(likelihood)\ information of the model, with the information from the prior
distribution.

Our motivation for the current paper involves the meta-analysis of 71 effect
sizes, which are heritability estimates that were reported by a collection
of behavioral-genetic studies of the heritability of antisocial behavior
(Talbott et al., 2012). Figure 1 presents a simple kernel probability
density estimate of the effect sizes. This estimate, which is skewed and has
at least two modes, clearly exhibits non-normality in the effect size
distribution. According to the standard Anderson-Darling test of normality
(Anderson \&\ Darling, 1952), the data rejects the null hypothesis that the
effect sizes are non-normal, at a .05 level of significance.

\begin{center}
\noindent --- Figure 1 ---
\end{center}

Arguably, a normal model (fixed-effects or random-effects) can provide an
adequate analysis of these data, specifically for the purposes of estimating
the overall mean of the effect size, and possibly also the overall variance.
However, such a model may not be adequate for predictive purposes, given the
lack of normality in the data. The importance of predicting from
meta-analytic data has been detailed in Higgins et \ al. (2009). As they
state:\ "Predictions are one of the most important outcomes of a
meta-analysis, since the purpose of reviewing research is generally to put
knowledge gained into future application. Predictions also offer a
convenient format for expressing the full uncertainty around inferences,
since both magnitude and consistency of effects may be considered." Accurate
prediction with non-normal data requires flexible models that support a
wider range of distributions, beyond the normal distribution. This has
already been recognized by Burr and\ Doss (2005) and Branscum and\ Hanson
(2008), who proposed Bayesian nonparametric models for meta-analysis. More
generally speaking, Bayesian nonparametric models are "nonparametric" in the
sense that they avoid the more restrictive assumptions of "parametric"
models, namely, that the data distribution can be fully-described by a few
finite number of parameters. For example, a normal model assumes that the
data distribution is unimodal and symmetric, and therefore can be entirely
described by a mean parameter and a variance parameter. A Bayesian
nonparametric model assigns a prior distribution to an infinite number (or a
very large number) of model parameters. This is done for the purposes of
defining a very flexible model that describes a wide range of data
distributions, including unimodal distributions, and more multimodal
distributions (M\"{u}ller \&\ Quintana, 2004).

Consequently, we propose a Bayesian nonparametric model for meta-analysis,
which is more flexible than the normal, fixed-effects and random-effects
models. Our model is a special case of the general model introduced by
Karabatsos and\ Walker (2012), which was studied in more general regression
settings. The new model specifies the effect-size distribution by an
infinite random-intercept mixture of normal distributions, conditional on
any covariate(s) of interest, with covariate-dependent mixture weights.
Therefore, the model is flexible enough to describe a very wide range of
effect-size distributions, including all normal distributions, as well as
all (smooth)\ distributions that are more skewed and/or multimodal. Also,
the model avoids the empirically-falsifiable assumption that effect-sizes
arise strictly from symmetric unimodal distributions, such as normal
distributions. Furthermore, the model's high flexibility encourages a rich
and graphical inference of the whole effect-size distribution, as previously
recommended for meta-analytic practice (Higgins et al., 2009).

Also, in the spirit of meta-regression analysis, our Bayesian nonparametric
meta-analysis model allows the whole effect-size distribution to change
flexibly and non-linearly as a function of key study-level covariates. This
feature permits a rich and flexible meta-regression analysis, whereas the
previous Bayesian nonparametric regression models do not account for
covariate information (see Burr \&\ Doss, 2005; Branscum \&\ Hanson, 2008).
Moreover, for a given meta-analysis, our Bayesian model can automatically
identify the subset of covariates that significantly predict changes in the
mean effect-size, in a model-based and non-ad-hoc fashion. Specifically, the
model makes use of spike-and-slab priors for the regression coefficients
that allow for automatic covariate (predictor)\ selection in the posterior
distribution. Such priors were developed for Bayesian normal linear
regression models (George \&\ McCulloch, 1997). Moreover, under either a
normal fixed-effects or normal random-effects model inferred under a
non-Bayesian (Frequentist)\ framework of maximum-likelihood estimation, the
identification or selection of significant study-level covariates
(predictors) is challenging because it deals with the standard issues of
multiple hypothesis testing over predictors (e.g., Thompson \&\ Higgins,
2002). The often-used stepwise procedures of covariate selection are known
to be ad-hoc and sub-optimal.

We now describe the layout of the rest of the paper. Since the paper covers
various key statistical concepts, it is necessary to first give them a brief
review in Section 2. In Section 2.1, we review the basic data framework of
meta-analysis, including effect sizes. In Section 2.2 we review of the
conventional normal fixed-effects and normal random-effects models, and in
Section 2.3 we briefly review the Bayesian statistical inference framework.
In Section 2.4 we review the traditional Bayesian meta-analytic models,
including conventional and more modern versions of normal fixed-effects and
normal random-effects models. In Section 3, we describe our new Bayesian
meta-analysis model. In Section 4, we review a standard criterion for
comparing the predictive performance between different Bayesian models that
are fit to a common data set, for the purposes of identifying the single
model that has best predictive-fit, i.e., of identifying the single model
that best describes the underlying population distribution of the sample
data. In Section 5, illustrates our Bayesian meta-analytic model through the
analysis of the large meta-analytic data set of behavioral genetic studies,
which was briefly described above, and which involves 24 covariates. In that
section, we also compare the predictive accuracy of our Bayesian
nonparametric model, against the predictive accuracy of conventional and
more modern normal fixed-effects and normal random-effects models. Section 6
ends with conclusions.\noindent

\section{Review of Meta-Analytic Modeling Concepts}

Before we review the key concepts underlying the various approaches to
meta-analysis, we describe some notation that we use in the remainder of
this paper. Following the standard notation of statistics, $\sim $ will mean
"distributed as"; $\mathrm{n}(\cdot |\mu ,v)$ denotes the ("bell-shaped")
probability density function (p.d.f.)\ of the normal distribution having
mean and variance $(\mu ,v)$; the p.d.f. of the $n$-variate normal
distribution with mean vector $\boldsymbol{\mu }$ and (symmetric and
positive-definite) variance-covariance matrix $\mathbf{\Sigma }$ is denoted
by $\mathrm{n}_{n}(\cdot |\boldsymbol{\mu },\mathbf{\Sigma })$; the p.d.f.
of a gamma distribution with shape and rate parameters $(a,b)$\ is denoted
by $\mathrm{ga}(\cdot |a,b);$ and the the p.d.f. of a uniform distribution
with minimum and maximum parameters $(a,b)$\ is denoted by $\mathrm{un}%
(\cdot |a,b)$. Also, we denote a cumulative distribution function (c.d.f.)
by a capital letter, such as $G$. Finally, $\delta _{\theta }(\cdot )$
denotes the degenerate distribution that assigns probability 1 (full
support) to the number $\theta $.

\subsection{\textit{Data Framework of Meta-Analysis}\label{Section: Data
Framework}}

In a typical meta-analysis context, data are available on $n$ study reports,
indexed by $i=1,\ldots ,n$. Each study reports an effect-size $y_{i}$ of
interest, based on $n_{i}$ observations, and provides information on the
sampling variance of the effect size. Also, each study provides information
about $p$ study characteristics, which are described by $p$ covariates $%
\mathbf{x}_{i}=(1,x_{1i},\ldots ,x_{pi})^{\intercal }$, in addition to a
constant (1) term for future notational convenience. A full meta-analytic
data set is denoted by $\mathcal{D}_{n}=\{(y_{i},\mathbf{x}_{i},\widehat{%
\sigma }_{i}^{2})\}_{i=1}^{n}$. \noindent

%TCIMACRO{\TeXButton{B}{\begin{table}[H] \centering}}%
%BeginExpansion
\begin{table}[H] \centering%
%EndExpansion
\begin{tabular}{lcc}
\hline\hline
\textbf{Effect-size }$\text{\textbf{Description}}$ & \textbf{Effect-size (}$%
y_{i}$) & \textbf{Variance (}$\widehat{\sigma }_{i}^{2}$) \\ \hline\hline
$%
\begin{array}{c}
\text{Unbiased standardized mean} \\ 
\text{difference, two independent } \\ 
\text{groups (Hedges, 1981). \ \ \ \ \ \ }%
\end{array}%
$ & $%
\begin{array}{c}
\dfrac{\widehat{\mu }_{1i}-\widehat{\mu }_{2i}}{\sqrt{\tfrac{(n_{1i}-1)%
\widehat{\sigma }_{1i}+(n_{2i}-1)\widehat{\sigma }_{2i}}{n_{1i}+n_{2i}-2}}}%
c^{\ast }%
\end{array}%
$ & $%
\begin{array}{c}
\left( \tfrac{n_{1i}+n_{2i}}{n_{1i}n_{2i}}+\tfrac{y_{i}^{2}}{2(n_{1i}+n_{2i})%
}\right) c^{\ast }; \\ 
c^{\ast }=1-\tfrac{3}{4(n_{1i}+n_{2i}-2)-1}%
\end{array}%
$ \\ \hline
$%
\begin{array}{c}
\text{Fisher z transformation \ \ } \\ 
\text{of the correlation }\widehat{\rho }_{i}.\text{\ \ \ \ \ \ } \\ 
\end{array}%
$ & $\tfrac{1}{2}\log \dfrac{1+\widehat{\rho }_{i}}{1-\widehat{\rho }_{i}}$
& $\dfrac{1}{n_{i}+3}$ \\ \hline
$%
\begin{array}{c}
\text{Log odds ratio for two \ \ \ } \\ 
\text{binary (0-1) variables. \ \ \ } \\ 
\end{array}%
$ & $\log \left( \dfrac{n_{11i}/n_{10i}}{n_{01i}/n_{00i}}\right) $ & $\dfrac{%
1}{n_{11i}}+\dfrac{1}{n_{10i}}+\dfrac{1}{n_{01i}}+\dfrac{1}{n_{00i}}$ \\ 
\hline\hline
\end{tabular}%
\caption{Examples of effect size statistics, along with corresponding
variances.}\label{Effect Sizes}%
%TCIMACRO{\TeXButton{E}{\end{table}}}%
%BeginExpansion
\end{table}%
%EndExpansion

Table \ref{Effect Sizes} presents some typical examples of effect-size
statistics that are often used in meta-analysis, along with their sampling
variances ($\widehat{\sigma }_{i}^{2}$) (Konstantopoulos, 2007; Borenstein,
2009, Fleiss \&\ Berlin, 2009). More generally, we may consider a sampling
covariance matrix for the $n$ study reports, $\widehat{\mathbf{\Sigma }}%
_{n}=(\widehat{\sigma }_{il})_{n\times n}$, having diagonal elements $%
\widehat{\sigma }_{ii}=\widehat{\sigma }_{i}^{2}$, where each off diagonal
element is the sampling covariance for a given effect-size pair $%
(y_{k},y_{l})$, with $k\neq l$ (Gleser \&\ Olkin, 2009). To maintain
notational simplicity throughout the paper, we will present the
meta-analytic models under the common assumption that $\widehat{\mathbf{%
\Sigma }}_{n}$ is a diagonal matrix (i.e., $\widehat{\mathbf{\Sigma }}_{n}=%
\mathrm{diag}(\sigma _{1}^{2},\ldots ,\sigma _{n}^{2})$), implying the
assumption of zero sampling covariances. Though, as we discuss in\ Section
3, this diagonal matrix assumption can be made for the Bayesian
nonparametric meta-analytic model, without loss of generality in terms of
being able to model covariances between distinct pairs of study effect size
reports.

\subsection{\textit{Traditional Meta-Analytic Models}\label{Section:
Traditional Meta Models}}

A traditional meta-analysis model assumes that, for a given set of data $%
\mathcal{D}_{n}=\{(y_{i},\mathbf{x}_{i},\widehat{\sigma }_{i}^{2})%
\}_{i=1}^{n}$, the effect-size distribution follows the general form: 
\begin{subequations}
\label{NREmeta}
\begin{align}
f(y_{i}|\mathbf{x}_{i},\widehat{\sigma }_{i}^{2};\boldsymbol{\zeta })& =%
\mathrm{n}(y_{i}|\mathbf{x}_{i}^{\mathbf{\intercal }}\boldsymbol{\beta }+\mu
_{0i}+\mu _{00t(i)},\widehat{\sigma }_{i}^{2}),\text{ \ }i=1,\ldots ,n;
\label{RElike} \\
\mathbf{x}_{i}^{\mathbf{\intercal }}\boldsymbol{\beta }& =\beta _{0}+\beta
_{1}x_{i1}+\cdots +\beta _{p}x_{ip};  \label{Line} \\
(\mu _{01},\ldots ,\mu _{0n})|\mathbf{\Sigma }_{0}& \sim \mathrm{n}_{n}(%
\mathbf{0},\sigma _{0}^{2}\mathbf{I}_{n}+\psi \mathbf{M}_{n});  \label{L2RE}
\\
\left. \mu _{00t}\right\vert \sigma _{00}^{2}& \sim \mathrm{n}(0,\sigma
_{00}^{2}),\text{ }t=1,\ldots ,T.  \label{L3RE}
\end{align}%
Typical meta-analytic models are special cases of the general normal model
shown in equation (\ref{NREmeta}). In all, the general traditional model (%
\ref{NREmeta}) has likelihood density $f(y|\mathbf{x},\widehat{\sigma }^{2};%
\boldsymbol{\zeta })$ with parameters $\boldsymbol{\zeta }=(\boldsymbol{%
\beta },\boldsymbol{\mu }_{0},\boldsymbol{\mu }_{00},\sigma _{0}^{2},\sigma
_{00}^{2},\psi )^{\intercal }$.\ These parameters are explained as follows.

The intercept parameter $\beta _{0}$ is interpreted as the \textit{mean
effect-size} over the given population of studies (Louis \&\ Zelterman,
1994). This interpretation holds true, provided that each of the $p$
covariates has data observations $(x_{k1},\ldots ,x_{kn})^{\intercal }$ that
have already been centered to have mean zero, as we assume throughout. Also,
the $p$ covariates are respectively parameterized by linear slope
coefficients ($\beta _{1},\ldots ,\beta _{p}$).

Also, $\boldsymbol{\mu }_{0}=(\mu _{01},\ldots ,\mu _{0n})^{\intercal }$ are
the level-2 random intercept parameters. Similarly, $\boldsymbol{\mu }%
_{00}=(\mu _{001},\ldots ,\mu _{00T})^{\intercal }$ are the \textit{level-3
random intercept parameters}, with each of the $n$ study reports being
nested within exactly one of $T\leq n$\ study reports ($t=1,\ldots ,T$), and
with $\mu _{00t(i)}$ meaning that the level-3 intercept $\mu _{00t}$ is
assigned to the $i$th study report. As shown in the model equations (\ref%
{L2RE}) and (\ref{L3RE}), the random intercepts $\boldsymbol{\mu }_{0}$ and $%
\boldsymbol{\mu }_{00}$ are each assumed to have a multivariate normal
distribution (probability density).

A normal fixed-effects model assumes that all the random intercept
parameters $(\boldsymbol{\mu }_{0},\boldsymbol{\mu }_{00})$ are zero. In
terms of the general normal meta-analytic model (\ref{NREmeta}), this
corresponds to the assumption of zero variances, i.e., $\sigma
_{0}^{2}=\sigma _{00}^{2}=\psi =0$ (Hedges \&\ Vevea, 1998). A 2-level
normal random-effects model allows for non-zero random intercepts $%
\boldsymbol{\mu }_{0}$, by allowing for nonzero variances $(\sigma
_{0}^{2},\psi ),$ as shown in equation (\ref{L2RE}). Typical normal
random-effects models assume that the random intercepts $\boldsymbol{\mu }%
_{0}$\ are uncorrelated, with $n$-variate normal density $\mathrm{n}_{n}(%
\boldsymbol{\mu }_{0}|\mathbf{0},\sigma _{0}^{2}\mathbf{I}_{n})$ (Hedges \&\
Vevea, 1998). However, it is possible to model correlations between the
level-2 random intercepts $\boldsymbol{\mu }_{0}$. For example, Stevens and
Taylor (2009) consider a 2-level normal random-effects model, which assumes
that the level-2 random intercepts have a $n$-variate normal density $%
\mathrm{n}_{n}(\boldsymbol{\mu }_{0}|\mathbf{0},\mathbf{\Sigma }_{0})$, with
a more general covariance structure $\mathbf{\Sigma }_{0}=\sigma _{0}^{2}%
\mathbf{I}_{n}+\psi \mathbf{M}_{n}$. Here, the parameter $\psi $ represents
the covariance between pairs of study reports. Also, $\mathbf{M}_{n}$ is a
fixed $n\times n$ indicator (0-1)\ matrix, with 1s specified in the
off-diagonal to reflect a-priori beliefs as to which pairs of the $n$ study
reports have correlated level-2 random intercepts, and with zeros specified
for all the other entries of $\mathbf{M}_{n}$. A 3-level normal
random-effects model allows for non-zero random intercepts $\boldsymbol{\mu }%
_{00}$, by allowing for a positive variance $\sigma _{00}^{2}$, as shown in
equation (\ref{L3RE}) of the general normal model (Konstantopoulos, 2011).

The general normal model described in equation (\ref{NREmeta}) can be
written explicitly as a normal mixture of multivariate normal model: 
\end{subequations}
\begin{equation*}
f(\mathbf{y}|\mathbf{X},\widehat{\mathbf{\Sigma }}_{n};\boldsymbol{\beta }%
,\sigma _{0}^{2},\psi ,\sigma _{00}^{2})=\dint \mathrm{n}_{n}(\mathbf{y}|%
\mathbf{X}\boldsymbol{\beta }+\boldsymbol{\mu }_{0}+\boldsymbol{\mu }%
_{00}^{\ast },\widehat{\mathbf{\Sigma }}_{n})\mathrm{d}G_{2}(\boldsymbol{\mu 
}_{0})\mathrm{d}G_{3}(\boldsymbol{\mu }_{00}),
\end{equation*}%
for effect-size data $\mathbf{y}=(y_{1},...,y_{n})^{\intercal }$, given the $%
n$-by-$(p+1)$ matrix $\mathbf{X}$ of row vectors $\mathbf{x}_{i}^{\intercal
} $ ($i=1,\ldots ,n$), $\boldsymbol{\mu }_{00}^{\ast }=(\mu _{00t(1)},\ldots
,\mu _{00t(n)})^{\intercal }$, and $\widehat{\mathbf{\Sigma }}_{n}=\mathrm{%
diag}(\sigma _{1}^{2},\ldots ,\sigma _{n}^{2}).$ Specifically, the mixture
model presented above assumes that the mixture distribution $G_{2}(%
\boldsymbol{\mu }_{0})$ is a multivariate distribution with probability
density $\mathrm{n}_{n}(\boldsymbol{\mu }_{0}|\mathbf{0},\sigma _{0}^{2}%
\mathbf{I}_{n}+\psi \mathbf{M}_{n})$, and that $G_{3}(\boldsymbol{\mu }%
_{00}) $ is a multivariate normal distribution with probability density $%
\mathrm{n}_{T}(\boldsymbol{\mu }_{00}|\mathbf{0},$ $\sigma _{00}^{2}\mathbf{I%
}_{n}).$

For any of the models described in this section, full maximum likelihood
methods can be used estimate the parameters $\boldsymbol{\zeta }=(%
\boldsymbol{\beta },\boldsymbol{\mu }_{0},\boldsymbol{\mu }_{00},\sigma
_{0}^{2},\sigma _{00}^{2},\psi )^{\intercal }$, from a given data set $%
\mathcal{D}_{n}$. Alternatively, the parameters of a random-effects model
can also be estimated by the restricted maximum-likelihood method, which
focuses estimation on the variance parameters (Harville, 1977; Raudenbush
\&\ Bryk, 2002, Ch. 3, 13-14; Stevens \& Taylor, 2009). For any one of the
individual models that is described in this section, the estimate of the
effect-size distribution of the underlying population is given by the
density $f_{n}(y|\mathbf{x}_{0},\widehat{\sigma }^{2};\widehat{\boldsymbol{%
\zeta }})=\mathrm{n}(y|\widehat{\beta }_{0},\widehat{\sigma }^{2}+\widehat{%
\sigma }_{0}^{2}+\widehat{\sigma }_{00}^{2})$ of the normal distribution,
given a maximum-likelihood estimate $\widehat{\boldsymbol{\zeta }}$ obtained
from a sample data set $\mathcal{D}_{n}$, and after controlling for all $p$
covariates via the covariate specification $\mathbf{x}=\mathbf{x}%
_{0}=(1,0,\ldots ,0)^{\intercal }$.

As an alternative to the maximum-likelihood approach, parameter estimation
can be performed in a fully-Bayesian framework (e.g., Higgins et al., 2009),
which is described next.

\subsection{\textit{Review of Bayesian Inference}}

For a general meta-analytic model, let $f(y|\mathbf{x},\widehat{\sigma }%
_{i}^{2};\boldsymbol{\zeta })$ denote the likelihood density of the
effect-size data point $y$, conditionally on covariates $\mathbf{x}$ and
model parameter $\boldsymbol{\zeta }$, which has space $\Omega _{\boldsymbol{%
\zeta }}$. A given meta-analytic data set $\mathcal{D}_{n}=\{(y_{i},\mathbf{x%
}_{i},\widehat{\sigma }_{i}^{2})\}_{i=1}^{n}$ has likelihood $L(\mathcal{D}%
_{n};\boldsymbol{\zeta })=\tprod\nolimits_{i=1}^{n}f(y_{i}|\mathbf{x}_{i},%
\widehat{\sigma }_{i}^{2};\boldsymbol{\zeta })$ under the model. In the
Bayesian approach to statistical inference, the model parameter $\boldsymbol{%
\zeta }$ is assigned a prior probability density\textit{\ }$\pi (\boldsymbol{%
\zeta })$ on the parameter space $\Omega _{\boldsymbol{\zeta }}$, and this
density reflects pre-experimental beliefs about the plausible values of the
parameter, for the meta-analytic data set $\mathcal{D}_{n}$ at hand. Then
according to Bayes' theorem, the data $\mathcal{D}_{n}$, via the model's
likelihood $L(\mathcal{D}_{n};\boldsymbol{\zeta })$, combines with the prior
density $\pi (\boldsymbol{\zeta })$, to yield a posterior density for $%
\boldsymbol{\zeta }$, defined by:%
\begin{equation}
\pi (\boldsymbol{\zeta }|\mathcal{D}_{n})=\frac{L(\mathcal{D}_{n};%
\boldsymbol{\zeta })\pi (\boldsymbol{\zeta })}{\tint_{\Omega _{\boldsymbol{%
\zeta }}}L(\mathcal{D}_{n};\boldsymbol{\zeta })\mathrm{d}\Pi (\boldsymbol{%
\zeta })}=\frac{\tprod\nolimits_{i=1}^{n}f(y_{i}|\mathbf{x}_{i},\widehat{%
\sigma }_{i}^{2};\boldsymbol{\zeta })\pi (\boldsymbol{\zeta })}{%
\tint_{\Omega _{\boldsymbol{\zeta }}}\tprod\nolimits_{i=1}^{n}f(y_{i}|%
\mathbf{x}_{i},\widehat{\sigma }_{i}^{2};\boldsymbol{\zeta })\mathrm{d}\Pi (%
\boldsymbol{\zeta })},  \label{Posterior}
\end{equation}%
where $\Pi (\boldsymbol{\zeta })$ denotes the c.d.f. of the prior density $%
\pi (\boldsymbol{\zeta })$. The posterior density describes the plausible
values of the model parameters $\boldsymbol{\zeta }$, given prior beliefs
about $\boldsymbol{\zeta }$ and data $\mathcal{D}_{n}$.

The specification of the prior density $\pi (\boldsymbol{\zeta })$ is an
important step in a Bayesian analysis, and the posterior $\pi (\boldsymbol{%
\zeta }|\mathcal{D}_{n})$ can be quite sensitive to the choice of prior,
especially when the sample size ($n$)\ is not large. Also, often in
practice, there is a lack of prior information about the parameters $%
\boldsymbol{\zeta }$ of the given model. This lack of prior information is
reflected by a "diffuse"\ prior probability density $\pi (\boldsymbol{\zeta }%
)$ that has high variance, and assigns rather-equal but broad support over
the parameter space $\Omega _{\boldsymbol{\zeta }}$. Such priors are often
referred to as "non-informative", even though technically speaking, a prior
cannot be fully non-informative. As a consequence of specifying a diffuse
prior $\pi (\boldsymbol{\zeta })$, the posterior density $\pi (\boldsymbol{%
\zeta }|\mathcal{D}_{n})$ becomes mostly determined by the data $\mathcal{D}%
_{n}$ likelihood $L(\mathcal{D}_{n};\boldsymbol{\zeta }),$ relative to the
prior.

Prediction is a basic function of statistical modeling. A Bayesian
meta-analytic model makes predictions of $Y$, given a chosen $\mathbf{x}$,
on the basis of the posterior predictive density:%
\begin{equation}
f_{n}(y|\mathbf{x},\widehat{\sigma }^{2})=\int f(y|\mathbf{x},\widehat{%
\sigma }^{2};\boldsymbol{\zeta })\pi (\boldsymbol{\zeta }|\mathcal{D}_{n})%
\mathrm{d}\boldsymbol{\zeta },  \label{ppd}
\end{equation}%
and this density has posterior predictive mean (expectation)\ $\mathrm{E}%
_{n}(Y|\mathbf{x},\widehat{\sigma }^{2})=\tint yf_{n}(y|\mathbf{x},\widehat{%
\sigma }^{2})\mathrm{d}y,$ and posterior predictive variance $\mathrm{Var}%
_{n}(Y|\mathbf{x},\widehat{\sigma }^{2})=\tint \{y-\mathrm{E}_{n}(Y|\mathbf{x%
},\widehat{\sigma }^{2})\}^{2}f_{n}(y|\mathbf{x,}\widehat{\sigma }^{2})%
\mathrm{d}y.$The posterior predictive density $f_{n}(y|\mathbf{x},\widehat{%
\sigma }^{2})$ provides an estimate of the true effect-size density
(distribution) for the underlying study population, given sample data $%
\mathcal{D}_{n}$ and covariates $\mathbf{x}$ of interest, under
squared-error loss (Aitchison, 1975).

In most Bayesian meta-analytic models, $\boldsymbol{\zeta }$ is a
high-dimensional parameter vector, and then the direct evaluation of the
posterior equations (\ref{Posterior}) and (\ref{ppd}) require prohibitive
high-dimensional integrations. In such situations, one may use Markov chain
Monte Carlo (MCMC) sampling methods to estimate such posterior densities of
the given model. Such methods include the Gibbs sampler (Gelfand \&\ Smith,
1990), Metropolis-Hastings algorithm (Chib \&\ Greenberg, 1995), and other
sampling algorithms (see for e.g., Brooks, Gelman, Jones, \& Meng, 2011).

\subsection{\textit{Review of Bayesian Normal Meta-Analytic Models} \label%
{Section: Traditional Bayes Meta-Analysis}}

In relation to the Bayesian inference framework described in Section 2.3,
consider the general normal meta-analytic model (\ref{NREmeta}), which has
likelihood density given by (\ref{RElike}), and which has parameters $%
\boldsymbol{\zeta }=(\boldsymbol{\beta },\boldsymbol{\gamma },\boldsymbol{%
\mu }_{0},\boldsymbol{\mu }_{00},\sigma _{0}^{2},\sigma _{00}^{2},\psi
)^{\intercal }$. Here, $\boldsymbol{\gamma }=(\gamma _{1},\ldots ,\gamma
_{p})^{\intercal }$ are included as parameters which respectively indicate
(0-1) whether or not the $p$ covariates are included ($\gamma =1$) or
excluded ($\gamma =0$) from the model. In typical practice involving such a
model (including special cases), the prior has the general form:%
\begin{equation}
\pi (\boldsymbol{\zeta })=\mathrm{n}(\boldsymbol{\beta }|\mathbf{0},\mathbf{V%
}_{\boldsymbol{\gamma }})\pi (\boldsymbol{\gamma })\mathrm{n}_{n}(%
\boldsymbol{\mu }_{0}|\mathbf{0},\sigma _{0}^{2}\mathbf{I}_{n}+\psi \mathbf{M%
}_{n})\mathrm{n}_{T}(\boldsymbol{\mu }_{00}|0,\sigma _{00}^{2}\mathbf{I}%
_{T})\pi (\sigma _{0}^{2},\psi )\pi (\sigma _{00}^{2}).  \label{NREprior}
\end{equation}%
In Bayesian meta-analytic modeling, it is common practice to specify a
diffuse normal prior density for the coefficients $\boldsymbol{\beta }$ by
taking $\mathbf{V}_{\boldsymbol{\gamma }}=v\mathbf{I}_{p+1}$, with $%
v\rightarrow \infty $ (e.g., DuMouchel \&\ Normand, 2000), with the implicit
assumption that $\pi (\boldsymbol{\gamma }=\boldsymbol{1})=1$.

Also, it is common in the practice of Bayesian random-effects modeling to
attempt to assign a non-informative prior for $(\sigma _{0}^{2},\sigma
_{00}^{2})$ via the specification of inverse-gamma priors $\pi (\sigma
_{0}^{2})=\mathrm{ga}(\sigma _{0}^{-2}|\epsilon ,\epsilon )$ and $\pi
(\sigma _{00}^{2})=\mathrm{ga}(\sigma _{00}^{-2}|\epsilon ,\epsilon )$, for
small choice of constant $\epsilon >0$ (Gelman, 2006), implying a prior
density of the form $\pi (\sigma _{0}^{2},\psi )=\pi (\sigma _{0}^{2})\delta
_{0}(\psi )$. Though, recall from Section \ref{Section: Traditional Meta
Models} that a more general multivariate normal $\mathrm{n}_{n}(\boldsymbol{%
\mu }_{0}|\mathbf{0},\sigma _{0}^{2}\mathbf{I}_{n}+\psi \mathbf{M}_{n})$
mixture distribution can be specified (Stevens \& Taylor, 2009). This
mixture distribution prior allows for correlated level-2 random intercepts $%
\boldsymbol{\mu }_{0}$ via a parameter $\psi $ that measures the covariance
between specific pairs of the total $n$ study reports, and where $\mathbf{M}%
_{n}=(m_{il})_{n\times n}$ is a fixed matrix which indicates (0-1) which
pairs of the study reports are expected to yield correlated level-2 random
intercepts $\boldsymbol{\mu }_{0}$, with zeros in the diagonal. For the
parameters $(\sigma _{0}^{2},\psi )$, Stevens and Taylor (2009) propose the
rather non-informative prior density%
\begin{equation}
\pi (\sigma _{0}^{2},\psi )=(c_{0}/(c_{0}+\sigma _{0}^{2})^{2})\mathrm{un}%
(\psi |-\sigma _{0}^{2}/(K-1),\sigma _{0}^{2}),
\label{Prior Dep L2 random effects}
\end{equation}%
where the first term in the product gives a log-logistic prior density for $%
\sigma _{0}^{2}$, where $c_{0}=\{n/\mathrm{tr}([\mathrm{diag}(\widehat{%
\mathbf{\Sigma }}_{n})]^{-1})\}^{1/2}$ is the harmonic mean of the sampling
variances $\widehat{\sigma }_{i}^{2}$ ($i=1,\ldots ,n$), and $%
K=\max_{i}\{\tsum\nolimits_{l=1}^{n}m_{il}\}$ is the largest group of
related study reports. This log-logistic prior is right-skewed,
highly-dispersed, with quartiles $(c_{0}/3,c_{0},3c_{0})$.

Simpler versions of the general normal random-effects model (\ref{NREmeta})
can be specified via appropriate straightforward modifications of the prior
density (\ref{NREprior}). A Bayesian 2-level normal random-effects model
which assumes $\sigma _{00}^{2}=0$ (i.e., $\boldsymbol{\mu }_{00}=\mathbf{0}$%
), but allows for correlated level-2 random intercepts $\boldsymbol{\mu }%
_{0} $ assigns the prior density $\pi (\sigma _{00}^{2})=\delta _{0}(\sigma
_{00}^{2});$ a 2-level normal random-effects model which assumes $\sigma
_{00}^{2}=\psi =0$ (i.e., $\boldsymbol{\mu }_{00}=\mathbf{0}$) and assume
independent level-2 random intercepts $\boldsymbol{\mu }_{0}$ assigns the
prior density $\pi (\sigma _{00}^{2},\psi )=\delta _{0}(\sigma
_{00}^{2})\delta _{0}(\psi )$; and a fixed-effects model which assumes $%
\sigma _{0}^{2}=\sigma _{00}^{2}=\psi =0$ (i.e., $\boldsymbol{\mu }_{0}=%
\mathbf{0}$ and $\boldsymbol{\mu }_{00}=\mathbf{0}$) assigns the prior
density $\pi (\sigma _{0}^{2},\sigma _{00}^{2},\psi )=\delta _{0}(\sigma
_{0}^{2})\delta _{0}(\sigma _{00}^{2})\delta _{0}(\psi )$.

Other simple modifications of the prior density (\ref{NREprior}) can be used
to specify other important versions of the general normal random-effects
model (\ref{NREmeta}), which consider different priors for the random
intercepts $(\boldsymbol{\mu }_{0},\boldsymbol{\mu }_{00})$. For example,
Gelman (2006) notes that posterior inference of the parameter $\sigma
_{0}^{-2}$, under the often-used "non-informative"\ gamma $\mathrm{ga}%
(\sigma _{0}^{-2}|\epsilon ,\epsilon )$ prior, is very sensitive to the
choice of small $\epsilon $, especially when the data support small values
of $\sigma _{0}^{2}$. Similarly for the gamma $\mathrm{ga}(\sigma
_{00}^{-2}|\epsilon ,\epsilon )$ prior for the parameter $\sigma _{00}^{2}$.
Therefore, in a situation where there is little prior information available
about the parameters $(\sigma _{0}^{2},\sigma _{00}^{2})$, he alternatively
recommends the specification of uniform prior densities $\pi (\sigma _{0})=$ 
\textrm{un}$(\sigma _{0}|0,b_{0})$ and $\pi (\sigma _{00})=$ \textrm{un}$%
(\sigma _{00}|0,b_{00})$ for reasonably-large values $(b_{0},b_{00})$,
whenever $n$ and $T$\ are both at least 5. When more prior information is
desired, say when $n$ is less than 5, he recommends the half-$t$ prior
density of the general form $\pi (\sigma _{0}^{2})\propto
(1+a_{0}^{-1}(\sigma _{0}/b_{0})^{2})^{-(a+1)/2}$, and similarly for the
level-3 variance parameter $\sigma _{00}$.

Finally, while it is common practice to assume a diffuse prior for the
regression coefficients $\boldsymbol{\beta }=(\beta _{0},\beta _{1},\ldots
,\beta _{p})^{\intercal }$, along with $\pi (\boldsymbol{\gamma }=%
\boldsymbol{1})=1$, in principle one may specify spike-and-slab priors for
the slope parameters $(\beta _{1},\ldots ,\beta _{p})$ in order to enable
automatic variable (covariate)\ selection via posterior inference (George
\&\ McCulloch, 1997), along with a diffuse prior $\beta _{0}\sim \mathrm{n}%
(0,v\rightarrow \infty ).$ These spike-and-slab priors are defined by
independent normal and Bernoulli prior densities, so that the $\mathrm{n}(%
\boldsymbol{\beta }|\mathbf{0},\mathbf{V}(\boldsymbol{\gamma }))\pi (%
\boldsymbol{\gamma })$ prior in (\ref{NREprior}) is based on: 
\begin{subequations}
\label{SpikeSlabPriors}
\begin{eqnarray}
\mathbf{V}_{\boldsymbol{\gamma }} &=&\mathrm{diag}(v\rightarrow \infty
,v_{0}(1-\gamma _{1})+v_{1}\gamma _{1},\ldots ,v_{0}(1-\gamma
_{p})+v_{1}\gamma _{p}) \\
\pi (\boldsymbol{\gamma }) &=&\dprod\nolimits_{k=1}^{p}\Pr (\gamma
_{k}=1)^{\gamma _{k}}[1-\Pr (\gamma _{k}=1)]^{1-\gamma _{k}},
\end{eqnarray}%
where $v_{0}$ is a small prior variance (e.g., $v_{0}=.001$), $v_{1}$ is a
large prior variance (e.g., $v_{0}=10$), and $\Pr (\gamma _{k}=1)$ is the
Bernoulli probability parameter that is often set to $.5$ in practice
(George \&\ McCulloch, 1997). So on the one hand, with prior probability $%
\Pr [\gamma _{k}=1]=.5$, the $k$th covariate is included in the model as a
"significant" predictor of the effect-size, by assigning its regression
coefficient $\beta _{k}$ a normal $\mathrm{n}(\beta _{k}|0,v_{1})$ prior
density that supports a large range of $\beta _{k}$ values. On the other
hand, with prior probability $\Pr (\gamma _{k}=0)=.5$, that covariate is
excluded from the model, by assigning its regression coefficient $\beta _{k}$
a normal $\mathrm{n}(\beta _{k}|0,v_{0})$ prior that places all its support
on values $\beta _{k}\approx 0$. Given MCMC\ samples from the posterior, $%
\pi (\boldsymbol{\zeta }|\mathcal{D}_{n})$, the $k$th covariate can be
viewed as a "significant predictor," when the posterior inclusion
probability of the covariate, $\Pr [\gamma _{k}=1|\mathcal{D}_{n}]$, is at
least .5 (Barbieri \&\ Berger, 2004). We assume throughout that the
spike-and-slab prior specifications are given by $v_{0}=.001$ and $v_{1}=10$%
. These specifications are consistent with the previous recommendation that
the ratio $v_{1}/v_{0}$ be no greater than 10,000, for the purposes of
reliably estimating the parameters $(\boldsymbol{\beta },\boldsymbol{\gamma }%
)$ via MCMC methods (George \&\ McCulloch, 1997, p. 368).

For any one of the models described in this section, the posterior
predictive density $f_{n}(y|\mathbf{x}_{0},\widehat{\sigma }^{2})$ provides
and estimate of the "overall"\ effect-size distribution of the underlying
population, and is a symmetric and unimodal distribution, conditionally on
covariates $\mathbf{x}=\mathbf{x}_{0}=(1,0,\ldots ,0)^{\intercal }$. For
example, under a Bayesian normal fixed-effects model, the posterior
predictive density $f_{n}(y|\mathbf{x}_{0},\widehat{\sigma }^{2})$ estimate
of the overall population effect-size distribution is a normal density
(distribution). Under a Bayesian normal 2-level random-effects model with $%
\mathrm{ga}(\sigma _{0}^{-2}|a,b)$ prior for independent level-2 random
intercepts $\boldsymbol{\mu }_{0}$, the posterior predictive estimate $%
f_{n}(y|\mathbf{x}_{0},\widehat{\sigma }^{2})$ of the population effect-size
distribution is given by a student density (distribution) (e.g., Denison et
al. 2002, Appendix). The student distribution is very similar to a normal
distribution, except that the student distribution has thicker tails.

For a given Bayesian normal meta-analytic model with prior density having
the general form (\ref{NREprior}), the posterior density $\pi (\boldsymbol{%
\beta },\boldsymbol{\gamma },\boldsymbol{\mu }_{0},\boldsymbol{\mu }%
_{00},\sigma _{0}^{2},\sigma _{00}^{2},\psi |\mathcal{D}_{n})$, the
posterior predictive density $f_{n}(y|\mathbf{x,}\widehat{\sigma }^{2})$,
and any functional of these densities, can estimated through the use of
standard MCMC Gibbs and Metropolis sampling algorithms for normal linear and
random-effects models (e.g., Gilks et al. 1993; Denison et al. 2002).

\section{The Bayesian Nonparametric \textbf{Meta-Analysis Model}}

For effect-size data, our Bayesian nonparametric meta-analysis model is
defined by an infinite random-intercepts mixture of regressions. The model
assumes the data likelihood: 
\end{subequations}
\begin{eqnarray}
f(y_{i}|\mathbf{x}_{i},\widehat{\sigma }_{i}^{2};\boldsymbol{\zeta })
&=&\dint \mathrm{n}(y_{i}|\mu _{0}+\mathbf{x}_{i}^{\intercal }\boldsymbol{%
\beta },\phi \widehat{\sigma }_{i}^{2})\mathrm{d}G_{\mathbf{x}}(\mu _{0})
\label{BNPmeta} \\
&=&\dsum\limits_{j=-\infty }^{\infty }\mathrm{n}(y_{i}|\mu _{0j}+\mathbf{x}%
_{i}^{\intercal }\boldsymbol{\beta },\phi \widehat{\sigma }_{i}^{2})\omega
_{j}(\mathbf{x}_{i}^{\intercal }\boldsymbol{\beta }_{\omega },\sigma
_{\omega }),\text{ \ }i=1,\ldots ,n.
\end{eqnarray}%
\newline
Using standard terminology for discrete mixture models (e.g., McLachlan \&\
Peel, 2000), $G_{\mathbf{x}}$ is the (discrete)\ mixing distribution which
depends on covariates $\mathbf{x}$; the component indices are given by $%
j=0,\pm 1,\pm 2,\ldots $, the component (kernel)\ probability densities are
given by the normal densities, $(\mathrm{n}(y_{i}|\mu _{0j}+\mathbf{x}%
_{i}^{\intercal }\boldsymbol{\beta },\phi \widehat{\sigma }%
_{i}^{2})_{j=-\infty }^{\infty }$; the component parameters are given by $%
(\mu _{0j})_{j=-\infty }^{\infty }$; and the mixing weights are given by $%
(\omega _{j}(\mathbf{x}_{i}^{\intercal }\boldsymbol{\beta }_{\omega },\sigma
_{\omega }))_{j=-\infty }^{\infty }$ which sum to 1 at every $\mathbf{x}\in 
\mathcal{X}$. Specifically, in the model, the mixture weights $\omega _{j}(%
\mathbf{x}_{i}^{\intercal }\boldsymbol{\beta }_{\omega },\sigma _{\omega })$
are each defined by the difference between two standard Normal(0,1)
cumulative distribution functions (c.d.f.s):%
\begin{equation*}
\omega _{j}(\mathbf{x}_{i}^{\intercal }\boldsymbol{\beta }_{\omega },\sigma
_{\omega })=\Phi (\{j-\mathbf{x}_{i}^{\intercal }\boldsymbol{\beta }_{\omega
}\}/\sigma _{\omega })-\Phi (\{j-1-\mathbf{x}_{i}^{\intercal }\boldsymbol{%
\beta }_{\omega }\}/\sigma _{\omega }).
\end{equation*}%
Thus, the mixture weights can be viewed as the categorical probabilities of
a cumulative-probits regression model (e.g.,\ McCullagh, 1980), for
infinitely-many ordered categories $j=0,\pm 1,\pm 2,\ldots $. Thus, it is
easy to see that the mixture weights $\omega _{j}(\mathbf{x}_{i}^{\intercal }%
\boldsymbol{\beta }_{\omega },\sigma _{\omega })$\ sum to 1 at every $%
\mathbf{x}\in \mathcal{X}$.\noindent \noindent\ Finally, our mixture model (%
\ref{BNPmeta}) provides a flexible (infinite) mixture of normal densities,
conditional on any covariates $\mathbf{x}$ of interest. The development of
our model is, in part motivated, by the well known fact that any smooth
probability density can be approximated arbitrarily-well by a suitable
discrete mixture of normal densities (e.g., Lo, 1984).

\begin{center}
\noindent --- Figure 2 ---
\end{center}

As shown in equation (\ref{BNPmeta}), for a set of data $\mathcal{D}%
_{n}=\{(y_{i},\mathbf{x}_{i},\widehat{\sigma }_{i}^{2})\}_{i=1}^{n}$, the
model assumes that each effect-size $y_{i}$ is distributed by according to a
probability density $f(y_{i}|\mathbf{x}_{i},\widehat{\sigma }_{i}^{2};%
\boldsymbol{\zeta })$ that is constructed by a mixture of an infinite number
of normal densities $\mathrm{n}(y_{i}|\mu _{0j}+\mathbf{x}_{i}^{\intercal }%
\boldsymbol{\beta },\phi \widehat{\sigma }_{i}^{2}),$ having corresponding
means $\mu _{0j}+\mathbf{x}_{i}^{\intercal }\boldsymbol{\beta }$ and mixture
weights $\omega _{j}(\mathbf{x}_{i}^{\intercal }\boldsymbol{\beta }_{\omega
},\sigma _{\omega })$ (for $j=0,\pm 1,\pm 2,\ldots $). Therefore, given
covariates $\mathbf{x}$ and model parameters $\boldsymbol{\zeta }$, the
model is flexible enough to allow the shape of the effect-size distribution
(density)\ $f(y|\mathbf{x},\widehat{\sigma }^{2};\boldsymbol{\zeta })$ to
take on virtually any form; this density can be unimodal symmetric, or
skewed, or more multimodal. Moreover, the model allows the entire shape and
location of the effect-size distribution\ (density)\ $f(y|\mathbf{x},%
\widehat{\sigma }^{2};\boldsymbol{\zeta })$ to change flexibly with the
covariates $\mathbf{x}$. The model has these flexibilities, because it
models the effect-size density $f(y|\mathbf{x},\widehat{\sigma }^{2};%
\boldsymbol{\zeta })$ by infinitely-many random intercept parameters $%
\boldsymbol{\mu }_{0}=(\mu _{0j})_{j=-\infty }^{\infty }$, corresponding to
infinitely-many covariate-dependent mixture weights $\{\omega _{j}(\mathbf{x}%
_{i}^{\intercal }\boldsymbol{\beta }_{\omega },\sigma _{\omega }):j=0,\pm
1,\pm 2,\ldots \}$.

In the Bayesian nonparametric model, the parameter $\sigma _{\omega }$
controls the level of multimodality of $f(y|\mathbf{x},\widehat{\sigma }^{2};%
\boldsymbol{\zeta })$. To explain, assume for the moment that $\mathbf{x}%
_{i}^{\intercal }\boldsymbol{\beta }=0$ and $f(y|\mathbf{x})=f(y|\mathbf{x},%
\widehat{\sigma }^{2}=1;\boldsymbol{\zeta })$, for simplicity, and with no
loss of generality. On the one hand, a small value of $\sigma _{\omega }$
indicates that $f(y|\mathbf{x})$ is unimodal, i.e., modeled as a unimodal
normal density $\mathrm{n}(y_{i}|\mu _{j},\widehat{\sigma }_{i}^{2})$ for a $%
j$ satisfying $j-1<\mathbf{x}^{\intercal }\boldsymbol{\beta }_{\omega }<j$,
with mixture weight $\omega _{j}(\mathbf{x}^{\intercal }\boldsymbol{\beta }%
_{\omega },\sigma _{\omega })$ near 1. This is because the function $\Phi (%
\mathbf{x}^{\intercal }\boldsymbol{\beta }_{\omega }/\sigma _{\omega })$ is
approximately 0 for $\mathbf{x}^{\intercal }\boldsymbol{\beta }_{\omega }<0$%
, while it is approximately 1 for $\mathbf{x}^{\intercal }\boldsymbol{\beta }%
_{\omega }>0$. As $\sigma _{\omega }$ approaches infinity, the mixture
weights become more spread out, and then $f(y|\mathbf{x})$ becomes
multimodal, with each mixture weight $\omega _{j}(\mathbf{x}^{\intercal }%
\boldsymbol{\beta }_{\omega },\sigma _{\omega })$ above zero and much less
than 1. These ideas are illustrated in Figure 2, which plots the mixture
weights and the corresponding density of our model, $f(y|\mathbf{x})$, for a
range of $\sigma _{\omega }$, given $\mathbf{x}^{\intercal }\boldsymbol{%
\beta }_{\omega }=.7$, given samples of $(\mu _{j},\sigma _{j}^{2})$ from a
normal-gamma distribution. As shown, the conditional density $f(y|\mathbf{x}%
) $ is unimodal when $\sigma _{\omega }$ is small, and $f(y|\mathbf{x})$
becomes more multimodal \noindent as $\sigma _{\omega }$ increases.\noindent%
\ The level of multimodality in the data is indicated by the posterior
distribution of $\sigma _{\omega }$, under the model.

The Bayesian nonparametric meta-analytic model (\ref{BNPmeta}) is completed
by the specification of a joint proper prior density $\pi (\boldsymbol{\zeta 
})$ for the infinitely-many model parameters $\boldsymbol{\zeta }=(%
\boldsymbol{\beta },$ $\boldsymbol{\gamma },$ $(\mu _{0j})_{j=-\infty
}^{\infty },$ $\phi ,$ $\boldsymbol{\beta }_{\omega },$ $\sigma _{\omega })$%
, according to the joint prior distributions: 
\begin{subequations}
\label{Priors}
\begin{eqnarray}
\mu _{0j}|\sigma _{0}^{2} &\sim &\mathrm{n}(\mu _{0j}|0,\sigma _{0}^{2}),%
\text{ \ }j=0,\pm 1,\pm 2,\ldots ;  \label{randInt} \\
\beta _{0} &\sim &\mathrm{n}(\beta _{0}|0,v\rightarrow \infty );
\label{Intercept prior} \\
(\beta _{k},\gamma _{k})|\phi &\sim &\mathrm{n}(\beta |0,\phi 10^{\gamma
_{k}}.001^{1-\gamma _{k}}).5^{\gamma _{k}}(1-.5)^{1-\gamma _{k}},\text{ \ }%
k=1,\ldots ,p;  \label{slope spike and slab} \\
\phi ^{-1} &\sim &\mathrm{ga}(\sigma _{\omega }^{-2}|a_{\phi }/2,a_{\phi }/2)
\label{DispPrior} \\
\sigma _{0}^{2} &\sim &\mathrm{un}(\sigma _{0}|0,b_{0})
\label{S20 BNP prior} \\
(\boldsymbol{\beta }_{\omega },\sigma _{\omega }) &\sim &\mathrm{n}_{p+1}(%
\boldsymbol{\beta }_{\omega }|\mathbf{0},\sigma _{\omega }^{2}10^{5}\mathbf{I%
}_{p+1})\mathrm{ga}(\sigma _{\omega }^{-2}|1,1).  \label{Mix weight prior}
\end{eqnarray}%
As shown, a diffuse prior is assigned to the overall mean effect-size
parameter $\beta _{0}$. Also, as shown in (\ref{slope spike and slab}), we
adapt the default spike-and-slab priors, to enable automatic covariate
(predictor) selection in the posterior distribution of our model (George \&
McCulloch, 1997). \noindent Also, the gamma prior for the inverse dispersion
parameter $\phi ^{-1}$ has mean $\mathrm{E}(\phi ^{-1})=1$ and variance $%
\mathrm{Var}(\phi ^{-1})=\allowbreak \frac{2}{a_{\phi }}$, with $a_{\phi }$
indicating the degree of `belief' in this prior (Nam, et al., 2003).
Furthermore, we specify uniform prior density \textrm{un}$(\sigma
_{0}|0,b_{0})$ for the variance $\sigma _{0}^{2}$ of the random intercepts $%
(\mu _{0j})_{j=-\infty }^{\infty }$, for a reasonably-large value $b_{0}$,
as consistent with previous recommendations (Gelman, 2006, Section 7.1).
Alternatively, one may specify a half-$t$ prior density for $\sigma _{0}$.
Most of the prior distributions in (\ref{Priors}) represent default and
rather diffuse choices of prior, which can be used in general meta-analytic
applications where prior information is typically limited. Of course, if for
a given meta-analytic data set, there is more prior (e.g., scientific)
information available about one or more of the model parameters, then the
prior distributions can be modified accordingly.

It is instructive to relate the parameters of the general normal
meta-analytic model of equation (\ref{NREmeta}) that are assigned a general
prior density of equation (\ref{NREprior}) (Sections \ref{Section:
Traditional Meta Models} and \ref{Section: Traditional Bayes Meta-Analysis}%
), with the parameters of the Bayesian nonparametric meta-analytic model.
Across both models, the linear regression coefficients $\boldsymbol{\beta }$%
, including the overall effect-size mean $\beta _{0}$, have the same
interpretation of how the mean effect-size depends on covariates; the
parameters $\boldsymbol{\gamma }=(\gamma _{1},\ldots ,\gamma
_{p})^{\intercal }$ have the same interpretation as (random)\ indicators of
which covariates are included as significant predictors of the models; the
infinitely-many random intercepts $(\mu _{0j})_{j=-\infty }^{\infty }$ of
the Bayesian nonparametric model have the same interpretation as the level-2
random intercepts $\boldsymbol{\mu }_{0}$ of a normal random-effects
meta-analytic model; and the parameter $\sigma _{0}^{2}$ has the same
interpretation as the variance of the level-2 random intercepts. A key
difference is that the Bayesian nonparametric model is a discrete mixture
model which specifies a covariate-dependent infinite-mixture distribution $%
G_{\mathbf{x}}$ for the random intercept parameter $\mu _{0}$, as opposed to
a hierarchical model or a random effects model. In contrast, a normal
random-effects model is a hierarchical model which specifies a normal
mixture distribution $G$ for the random intercept parameter, such that the
mixture distribution is not covariate dependent. Moreover, for the Bayesian
nonparametric model, the discrete mixture distribution $G_{\mathbf{x}}$
induces (random) clusterings among the $n$ study reports $y_{i}$ (via the
posterior distribution of the model), in terms of the random intercept
parameter $\mu _{0}$. This clustering feature of the Bayesian nonparametric
model enables the model to account for correlations among the study reports $%
y_{i}$ ($i=1,\ldots ,n$), lessening the need to specify a non-diagonal
sampling covariance matrix $\widehat{\mathbf{\Sigma }}_{n}$ to account for
correlated effect-sizes.

Following Bayes' theorem, the data likelihood $L(\mathcal{D}%
_{n})=\tprod\nolimits_{i=1}^{n}f(y_{i}|\mathbf{x}_{i},\widehat{\sigma }%
_{i}^{2};\boldsymbol{\zeta })$ updates the prior density $\pi (\boldsymbol{%
\zeta })$, to a posterior density $\pi (\boldsymbol{\zeta }|\mathcal{D}_{n})$%
, given by equation (\ref{Posterior}). Then the posterior predictive density
is given by equation (\ref{ppd}), which gives an estimator of the true
effect-size density in the study population, given data $\mathcal{D}_{n}$
and covariates $\mathbf{x}$ of interest. Also, recall that for the task of
covariate\ selection, the $k$th covariate can be viewed as a "significant
predictor," when the posterior inclusion probability of the covariate, $\Pr
[\gamma _{k}=1|\mathcal{D}_{n}]$, is at least .5 (Barbieri \&\ Berger,
2004). Finally, the level of multimodality in the density $f(y|\mathbf{x},%
\widehat{\sigma }^{2},\boldsymbol{\zeta })$ is indicated by the posterior
distribution of $\sigma _{\omega }$.

Karabatsos and Walker (2012) describe the Markov Chain Monte Carlo (MCMC)\
methods can be used to perform inference of the posterior density $\pi (%
\boldsymbol{\zeta }|\mathcal{D}_{n})$, of the posterior predictive density $%
f_{n}(y|\mathbf{x},\widehat{\sigma }^{2})$ of the model, and to perform
inference of any functional of these densities.

\section{Bayesian Predictive Model Assessment Methods}

Model selection is the practice of comparing different models that are
fitted to a common sample data set, and then identifying the single model
that best describes or predicts the underlying population distribution of
the sample data. In meta-analytic practice, it is often of interest to
perform model selection (e.g., Sutton, 2000, Section 11.7.3). For example,
model selection is used in meta-analysis to choose between the fixed-effects
and random-effects model (Borenstein et al. 2010), or to select important
predictors of the effect-size\ in a regression setting (Higgins \&\
Thompson, 2004).

After $M$\ meta-analytic models are fit to a data set, $\mathcal{D}_{n}$,
the predictive performance of each Bayesian model $m\in \{1,\ldots ,M\}$ can
be assessed by the mean-square posterior predictive-error criterion 
\end{subequations}
\begin{eqnarray}
D(m) &=&\sum_{i=1}^{n}\{y_{i}-\text{\textrm{E}}_{n}(Y_{i}|\mathbf{x}_{i},%
\widehat{\sigma }_{i}^{2},m)\}^{2}+\sum_{i=1}^{n}\mathrm{Var}_{n}(Y_{i}|%
\mathbf{x}_{i},\widehat{\sigma }_{i}^{2},m)  \label{Pred Criterion} \\
&=&\sum_{i=1}^{n}\dint (y-y_{i})^{2}f_{n}(y|\mathbf{x}_{i},\widehat{\sigma }%
_{i}^{2},m)\mathrm{d}y=\sum_{i=1}^{n}D_{i}(m)
\end{eqnarray}%
(Laud \&\ Ibrahim, 1995; Gelfand \& Ghosh, 1998). The criterion (\ref{Pred
Criterion}) is a standard criterion that is often used for the assessment
and comparison of Bayesian models (e.g., Gelfand \& Banerjee, 2010). Among
the $M$\ Bayesian meta-analytic models that are compared, the model with the
smallest value of $D(m)$ is identified as the one that best describes the
underlying population distribution of the given sample data set $\mathcal{D}%
_{n}$. The first term of (\ref{Pred Criterion}) measures data
goodness-of-fit, and the second term is a penalty that is large for models
which either over-fit or under-fit the data, as in other classical model
selection criteria. Taking the square root, $\surd D(m)$, makes the
criterion interpretable on the original scale of the effect size ($y$).
Similarly, the individual square-root quantities $\surd D_{i}(m)$ (for $%
i=1,\ldots ,n$) can provide a detailed assessment about a model's predictive
performance. A large value of $\surd D_{i}(m)$ would indicate that the
observed effect-size $y_{i}$ is an outlier under the model.

\section{\textbf{Illustration\label{Section: Illustration}}}

In this section, we illustrate all the methods that we presented in Sections
2-4, through the meta-analysis of a large real data set involving 24
covariates. In this data analysis, we use the $D(m)$ predictive mean-square
error criterion to compare the predictive accuracy of the Bayesian
nonparametric meta-analysis model, and the various Bayesian normal
fixed-effects and normal random-effects models. We also compare some of the
parameter estimates between these Bayesian models, as well as the parameter
estimates of normal fixed-effects and random-effects models estimated either
under full maximum likelihood (MLE) or restricted maximum likelihood (REML).
For each data analysis, each covariate was previously z-standardized to have
mean 0 and variance 1, by taking $x_{ki}=(x_{ki}^{\prime }-\widehat{\mu }%
_{k}^{\prime })/\widehat{\sigma }_{X(k)}^{\prime }$ for $i=1,\ldots ,n$,
given the mean and standard deviation $(\widehat{\mu }_{k}^{\prime },%
\widehat{\sigma }_{X(k)}^{\prime })$ of the original covariate data $%
(x_{k1}^{\prime },\ldots ,x_{kn}^{\prime })$. Then the the estimate of the
intercept parameter ($\beta _{0}$) is interpretable as the mean study
effect-size, and the $\boldsymbol{\beta }$ coefficients are all
interpretable on a common scale.

In total, we will consider a total of 16 Bayesian meta-analytic models for
the data set, including the Bayesian nonparametric model, along with various
fixed effects models and various 2-level or 3-level normal-effects models,
which among other things, differ as to whether or not they have covariates,
whether or not they have spike-and-slab priors for covariate selection. For
all of these models, we specified the same prior densities for parameters
that the models shared in common, in order to place the Bayesian model
comparisons on a rather equal-footing. These priors were generally
consistent with the recommendations of the previous literature (see Sections %
\ref{Section: Traditional Bayes Meta-Analysis}, 3). Specifically, for all
Bayesian models, as follows:\ we assigned the normal prior density $\pi
(\beta _{0})=\mathrm{n}(\beta _{0}|0,v\rightarrow \infty )$ \ (with $%
v=10^{5} $); we assigned diffuse normal priors $\pi (\beta _{k})=\mathrm{n}%
(\beta _{k}|0,v\rightarrow \infty )$ \ (with $v=10^{5}$), $k=1,\ldots ,p=24,$
to the slope coefficients of all models with covariates and without
spike-and-slab priors; we assigned hyper-prior variances $v_{0}=.001$ and $%
v_{1}=10$, and Bernoulli prior parameters $\Pr (\gamma _{k}=1)=\tfrac{1}{2}$
($k=1,\ldots ,p=24$), for all models with covariates and spike-and-slab
priors (in terms of equation \ref{SpikeSlabPriors}); we assigned the uniform
prior density $\pi (\sigma _{0})=\mathrm{un}(\sigma _{0}|0,100)$ with large
scale (100) for the variance parameter of the Bayesian nonparametric model
and all normal 2-level random effects models; we specified the rather
non-informative prior for $(\sigma _{0}^{2},\psi )$ (equation (\ref{Prior
Dep L2 random effects}), Section \ref{Section: Traditional Bayes
Meta-Analysis}), for all Bayesian 2-level normal random-effects models that
allow for correlated random intercepts (Stevens \& Taylor, 2009); and we
specified the uniform prior density $\pi (\sigma _{00})=\mathrm{un}(\sigma
_{00}|0,100)$ with large scale (100) for the level-3 variance parameter $%
\sigma _{00}^{2},$ for all 3-level normal random-effects models. For the
Bayesian nonparametric model in particular, we also specified rather diffuse
(high-variance) priors $\phi ^{-1}\sim \mathrm{ga}(\sigma _{\omega
}^{-2}|.5/2,.5/2)$, $\boldsymbol{\beta }_{\omega }|\sigma _{\omega }^{2}\sim 
\mathrm{n}_{p+1}(\boldsymbol{\beta }_{\omega }|\mathbf{0},\sigma _{\omega
}^{2}10^{5}\mathbf{I}_{p+1})$, and $\sigma _{\omega }^{-2}\sim \mathrm{ga}%
(\sigma _{\omega }^{-2}|1,1)$ (see equations (\ref{DispPrior}) and (\ref{Mix
weight prior}), Section 3).

For the application of each Bayesian model, the posterior distribution of
the parameters was estimated on the basis of 200,000 MCMC samples, after
trace plots indicated that MCMC\ samples of key model parameters and of the $%
D(m)$ criterion stabilized and mixed well, and after 95\%\ Monte Carlo
confidence intervals (MCCIs)\ of these quantities attained half-widths that
were small for practical purposes, i.e., that typically ranged between .01
and .05, and not exceeding .1. Again, these procedures accord with previous
recommendations for checking the quality of MCMC-based posterior estimates,
based on a single MCMC\ run (Geyer, 1992; 2011, Chapter 1). Our Bayesian
nonparametric meta-analytic model was estimated using a program code we
wrote in the MATLAB (Natick, VA) software language, for an earlier paper
(Karabatsos \&\ Walker, 2012). The code for the meta-analytic model is
provided along with this paper. The Bayesian normal fixed-effects and
random-effects models were each estimated using code we wrote in MATLAB.
Finally, each of the twelve 2-level and 3-level random-effects models,
assuming uncorrelated random intercepts, were also estimated by full maximum
likelihood and restricted maximum likelihood, using the nmle package
(Pinheiro, et al. 2010) of the R\ statistical software (R Development Core
Team, 2012).

\subsection{\noindent \textit{Behavioral Genetics Data}{\textbf{\protect%
\medskip }}}

Antisocial behavior, which includes aggression, willingness to violate rules
and laws, defiance of adult authority, and violation of social norms (Walker
et al. 2003), is the most frequent reason why children are referred for
mental health services in schools (Adelman \& Taylor, 2010). Yet, it is the
most intractable of all behavior and mental health problems, is challenging
to treat, and must be addressed across the lifespan (Moffitt, 1993;
2005).\noindent

\begin{center}
\noindent --- Figure 3 ---
\end{center}

To advance understanding and treatment, many behavioral genetic studies have
investigated the heritability of antisocial behavior, by correlating ratings
of antisocial behavior among monozygotic (MZ) identical twin pairs, and
among dizygotic (DZ) fraternal\ twin pairs. In each study, the ratings were
done either by the mother, father, teacher, self, or an observer. Then the
heritability, defined as the proportion of phenotypic variance explained by
genetic factors, is estimated by twice the difference between the MZ
correlation and DZ correlation, for twins of the same sex. Specifically, for
a given gender, suppose that $n_{MZ}$ monozygotic (MZ) identical\ twin pairs
yield a correlation $\widehat{\rho }_{MZ}$ on an antisocial behavior trait,
such as conduct disorder, aggression, delinquency, and externalizing
behavior. Also, suppose that $n_{DZ}$ dizygotic (DZ) fraternal\ twin pairs
yield a correlation $\widehat{\rho }_{DZ}$ on the same trait. Then the
heritability of the antisocial behavior trait is estimated by:%
\begin{equation}
\widehat{h}^{2}=2(\widehat{\rho }_{MZ}-\widehat{\rho }_{DZ})  \label{Heret}
\end{equation}%
(Falconer \&\ Mackay, 1996). This effect-size statistic (\ref{Heret}) has
sampling variance:%
\begin{equation*}
\widehat{\sigma }^{2}=4[\{(1-\widehat{\rho }_{MZ}^{2})^{2}/n_{MZ}\}+\{(1-%
\widehat{\rho }_{DZ}^{2})^{2}/n_{DZ}\}].
\end{equation*}

We identified 29 independent studies that provided the information necessary
to estimate antisocial behavior heritability, for the $n=71$ independent
samples (study reports) of MZ-DZ\ twin comparisons (Talbott, et al. 2012).
These studies were published during years 1966 through 2009, and their full
references are listed in the Appendix. There were 2-3 heritability estimates
per study on average, and each study provided between 1 to 10
estimates.\noindent\ The left panel of Figure 3 presents the heritability
estimates of antisocial behavior (i.e., the effect-size observations),
stratified by gender, by rater type, and by the studies which were
numerically identified by publication year order (see Appendix). The one
slightly-negative estimate ($-.06$) may have resulted from sampling error
(Gill \&\ Jensen, 1968), as suggested by its relatively-large
variance.\noindent

Here, it is of interest to perform a meta-analysis of the studies, to learn
about the overall heritability (effect-size) distribution for the underlying
study population, as well as to learn how heritability changes with key
study-level covariates. Again, analyses will be performed using the various
fixed-effects and random-effects models under maximum likelihood estimation,
and using our Bayesian nonparametric model. A total of 24 covariates were
identified. They include publication year, the square root of the
heritability variance (denoted SE(ES)) to provide an investigation of
publication bias; indicators (0-1) of female status (49\%) versus male; ten
indicators of antisocial behavior ratings done by mother (mean=.53), father
(.06), teacher (.24), self (.15), independent observer (.04), and ratings
done on conduct disorder (.03), aggression (.40), delinquency (.10), and
externalizing (.48) antisocial behavior; an indicator of whether a weighted
average of heritability measures was taken within study over different
groups of raters who rated the same twins (28\% of cases, for which the ten
indicator covariates are scored as group proportions), mean age of the study
subjects in months\ (overall mean=119.8, s.d.=49.5); indicators of
hi-majority ($\geq $ 60\%) white twins in study (94\%), zygosity obtained by
questionnaire (80\%), or through DNA samples (68\%), study inclusion of low
socioeconomic (SES) status level subjects (20\%) and mid-to-high SES
subjects (90\%), missing SES information (10\%), representative sample
(85\%), longitudinal sample (85\%), and location of the study in terms of
latitude and longitude.\noindent \noindent

\noindent \noindent 
%TCIMACRO{\TeXButton{B}{\begin{table}[H] \centering}}%
%BeginExpansion
\begin{table}[H] \centering%
%EndExpansion
\begin{tabular}{lcclc}
\hline
Model & $D(m)$ &  & Model & $D(m)$ \\ \hline
BNP-ss & $0.6$ &  & D2L-x & $5.5$ \\ 
D2L-0 & $4.8$ &  & 3L-x & $5.5$ \\ 
2L-0, by MZ-DZ & $4.8$ &  & 2L-0, by Study & $5.8$ \\ 
3L-0 & $4.8$ &  & FE-0 & $5.9$ \\ 
D2L-ss & $5.4$ &  & 2L-ss, by Study & $6.0$ \\ 
2L-ss, by MZ-DZ & $5.4$ &  & FE-ss & $6.0$ \\ 
3L-ss & $5.4$ &  & 2L-x, by Study & $6.0$ \\ 
2L-x, by MZ-DZ & $5.5$ &  & FE-x & $6.0$ \\ \hline
\end{tabular}%
%TCIMACRO{%
%\TeXButton{caption}{\caption{For the behavioral genetics data, a comparison of the predictive mean square error criterion, between the Bayesian nonparametric model, and various normal fixed-effects (FE) models, and various normal 2-level (2L), dependent 2-level (D2L), and 3-level random effects (RE) models, each of which either has no covariates (0), or 24 covariates (x), or 24 covariates with stochastic search variable selection (ss). Also, each 2L or D2L model groups the random intercept parameters by the 29 studies (by Study), or groups them by the 71 independent samples of twins (by MZ-DZ).}}}%
%BeginExpansion
\caption{For the behavioral genetics data, a comparison of the predictive mean square error criterion, between the Bayesian nonparametric model, and various normal fixed-effects (FE) models, and various normal 2-level (2L), dependent 2-level (D2L), and 3-level random effects (RE) models, each of which either has no covariates (0), or 24 covariates (x), or 24 covariates with stochastic search variable selection (ss). Also, each 2L or D2L model groups the random intercept parameters by the 29 studies (by Study), or groups them by the 71 independent samples of twins (by MZ-DZ).}%
%EndExpansion
\label{Gene Results}%
%TCIMACRO{\TeXButton{E}{\end{table}}}%
%BeginExpansion
\end{table}%
%EndExpansion

Given the structure of the data, with the $n=71$ heritability (effect-size)\
estimate reports nested within the 29 studies, any one of at least 15 normal
fixed-effects models or normal random-effects models can be considered for
the purposes of meta-analysis. Specifically, they include: fixed-effects
models; 2-level random-effects models, with level-2 random intercepts $%
\boldsymbol{\mu }_{0}$ assumed to be independent via the specification of a
multivariate normal $\mathrm{n}_{n}(\boldsymbol{\mu }_{0}|\mathbf{0},\sigma
_{0}^{2}\mathbf{I}_{n})$ distribution, with either a structure that has each
of the 71 independent samples of MZ-DZ\ twin comparisons defining its own
group, or a grouping structure has each of the 29 studies defining its own
group; 2-level random-effects models that allow for dependent level-2 random
intercepts via the specification of a multivariate normal $\mathrm{n}_{n}(%
\boldsymbol{\mu }_{0}|\mathbf{0},\sigma _{0}^{2}\mathbf{I}_{n}+\psi \mathbf{M%
}_{n})$ distribution (Stevens \&\ Taylor, 2009) with the binary (0-1) matrix 
$\mathbf{M}_{n}$ indicating which pairs of the 71 study reports belong in
the same study; 3-level random-effects models, with the 71 heritability
(effect-size)\ estimate reports (level 2)\ nested within the 29 studies
(level 2), and with random intercepts modeled respectively by the
multivariate normal densities $\mathrm{n}_{n}(\boldsymbol{\mu }_{0}|\mathbf{0%
},\sigma _{0}^{2}\mathbf{I}_{n})$ and $\mathrm{n}_{n}(\boldsymbol{\mu }_{00}|%
\mathbf{0},\sigma _{00}^{2}\mathbf{I}_{n})$; with each model having either
no covariates, or having all 24 covariates with no spike-and-slab priors for
automatic covariate selection, or having all 24 covariates along with
spike-and-slab priors for automatic covariate selection via the posterior
distribution. Finally, we also analyze the data set using the Bayesian
meta-analytic model, which includes all 24 covariates, and which assigns
spike-and-slab priors for automatic covariate selection via the posterior
distribution.

For the behavioral genetics data, Table \ref{Gene Results} compares the
estimate of the mean-squared predictive error criterion, $D(m)$, between the
15 Bayesian normal fixed-effects models and normal random-effects models,
and the Bayesian nonparametric meta-analytic model. Among all the models
compared, the Bayesian nonparametric model attained the smallest value of
the predictive mean-square error criterion $D(m)$, by a relatively-large
margin. Hence, among all the models compared, the Bayesian nonparametric
model provides the best description of the study-population effect-size
distribution that underlies the (sample) behavioral genetics data set. The
meta-analytic models that attained the second best value of the criterion
had about 8 times the mean squared predictive error, compared to our
Bayesian nonparametric model. Also, for the Bayesian nonparametric model,
the 5-number summary of the estimates of the predictive residuals $\surd
D_{i}(m)$ was $(.0,.0,.1,.1,.3)$, over the 71 heritability effect-size
observations $y_{i}$. So none of the observations appeared to be outliers
under the model.

The mean heritability (effect-size) estimate ($\widehat{\beta }_{0}$) was
quite similar among all the 16 Bayesian models, ranging from .49 to .51;
while the posterior mean estimates $(\widehat{\sigma }_{0}^{2},\widehat{%
\sigma }_{00}^{2},\widehat{\psi })$ of the random intercept variances ranged
between .00 to .02. Similar estimates were obtained from normal
fixed-effects and random-effects models that were estimated either under
maximum likelihood, or by restricted maximum likelihood. For all the
Bayesian models assigned spike-and-slab priors, the posterior inclusion
probabilities $\Pr [\gamma =1|\mathcal{D}_{71}]$ did not exceed .05 for all
24 covariates, well below the significance threshold of $.5$. For the
Bayesian nonparametric model, the marginal posterior mean (standard
deviation)\ estimate of the dispersion parameter $\phi $ was $.09$ ($.03$).

As mentioned, the Bayesian nonparametric model is an infinite mixture model
that is able to account for all possible shapes and locations of effect-size
distributions, including all normal distributions. Meanwhile, in terms of
the predictive mean-square error criterion $D(m)$, the Bayesian
nonparametric model far-outperformed all other normal fixed-effects and
normal random-effects models, which assume more strictly assume normal
effect-size densities. These facts together suggest that the data set
violates the assumptions of effect-size normality. For the Bayesian
nonparametric model, the right panel of Figure 4 presents the posterior
predictive estimate of the overall heritability (effect-size)\ distribution,
for the underlying population of studies. This estimate is conditional on
the covariates $\mathbf{x}=\mathbf{x}_{0}=(1,0,\ldots ,0)^{\intercal }$, and
also, it is conditioned on the minimum effect-size variance $\sigma _{i}^{2}$
of .0001 over all 71 heritability reports, so that this distribution
reflects information from a large-sample\ study. According to this estimate
of the overall heritability (effect-size)\ distribution, there is some
evidence of skewness ($-.1$), with the overall mean ($.50$) and median ($.51$%
) heritability (effect-size) being slightly different. Moreover, there seems
to be two modes in this distribution, one at about the mean of $.50$, and
the other at about $.35$, suggesting there are about two "significant"\
heritabilities (effect sizes) in the population, not only one. Upon closer
inspection, the modes appear to be at heritability values of .38 and .51. So
both modes can provide information that contribute to the accumulation of
evidence about the overall heritability (effect-size) for the substantive
researchers of behavioral genetics.

\begin{center}
\noindent --- Figure 4 ---
\end{center}

The first panel of Figure 4 shows the median (50\%ile)\ and interquartile
range (i.e., 25\%ile and 75\%ile) of the posterior predictive estimate of
the heritability (effect-size) distribution, by SE(ES) (and by corresponding
effect-size variance $\{$SE(ES)$\}^{2}=\widehat{\sigma }^{2}$). As shown,
the median effect-size has a rather nonlinear relationship with SE(ES), but
the lack of strong relationship of the median effect-size with SE(ES)
further confirms a lack of evidence of publication bias in the data.\noindent

Finally, another important issue in this area of behavioral genetics deals
with the issue of informant discrepancy; that is, the issue of whether the
heritability (effect-size) estimates are the same across raters, or whether
they depend on rater type, and further the heritability estimates may also
depend on ratee age. Recall that each of the 71 heritability (effect-size)
estimates were obtained from ratings of antisocial behavior that were made
by one of 5 rater types, namely, the mother, the father, the teacher, the
self, or an independent observer. To address this research issue in detail,
the remaining five panels of Figure 4 present the posterior predictive
estimates of the effect-size distributions, conditional on covariates $%
\mathbf{x}$ that indicate rater type and ratee age, while controlling for
all other non-constant covariates by setting them to zero, and while
conditioning on the effect-size variance estimate $\widehat{\sigma }%
_{i}^{2}=.0001$ that reflects a large-sample\ study. As shown in these
panels, the heritability-age correlation seems to be sightly negative for
all rater types. Moreover, the heritability distributions seemed to be
similar among mother, father, self, and independent observer raters, while
the teacher raters have noticeably different heritability distributions.

\section{\textbf{Discussion\label{Section: Discussion}}}

In this paper, we have proposed a Bayesian nonparametric model for
meta-analysis, and demonstrated its suitability for meta-analytic data sets
that give rise to asymmetric and more multimodal effect-size population
distributions. As mentioned, the traditional normal fixed- and
random-effects models, while frequently used for meta-analysis, are not
fully satisfactory because they make empirically-falsifiable assumptions
about the data. They include the assumption that the effect-sizes are
normally-distributed (conditionally on model parameters). As we have shown,
empirical violations of such an assumption can negatively affect the
accuracy of prediction of meta-analytic data.

In contrast to the traditional models, our proposed Bayesian nonparametric
model flexibly accounts for all distributions of the effect-sizes, including
all normal distributions. For the real data set that was analyzed in this
paper, this flexibility enabled the Bayesian model to provide a better
description of the underlying effect-size distribution of the underlying
study population. At the same time, the model provides a richer description
of meta-analytic data, by allowing the data analyst to infer the whole
distribution of effect-sizes, over studies, and to infer how the whole
distribution changes as a function of key study-level covariates. Thus, the
model goes beyond the mean as the measure of an overall effect-size.
Furthermore, for the given meta-analytic data set at hand, the Bayesian
nonparametric model automatically identifies important study-level
predictors of the mean effect-size.\bigskip \bigskip

\noindent {\textbf{Acknowledgements\medskip }}

This research is supported by grant SES-1156372, from the NSF\ program in
Methodology, Measurement, and Statistics. The author gives thanks to the
Editor, and to anonymous Associate Editor and two reviewers, for detailed
suggestions that have helped improve the presentation of this manuscript.

\bigskip

\noindent \textbf{References}

\begin{description}
\item Adelman, H. \& Taylor, L. 2010. \textit{Mental health in schools:
Engaging learners, preventing problems, improving schools}, Corwin Press,
Thousand Oaks, CA.

\item Aitchison, J. 1975. Goodness of prediction fit, \textit{Biometrika} 
\textbf{62}: 547-554.

\item Anderson, T. \& Darling, D. 1952. Asymptotic theory of certain
"goodness-of-fit" criteria based on stochastic processes, \textit{Annals of
Mathematical Statistics} \textbf{23}: 193-212.

\item Barbieri, M. \& Berger, J. 2004. Optimal predictive model selection, 
\textit{Annals of Statistics} \textbf{32}: 870-897.

\item Berkey, C., Hoaglin, D., Mosteller, F. \& Colditz, G. 1995. A
random-effects regression model for meta-analysis, \textit{Statistics in
Medicine }\textbf{14}: 395-411.

\item Borenstein, M. 2009. Effect sizes for continuous data, in H. Cooper,
L. Hedges \& J. Valentine (eds), \textit{The Handbook of Research Synthesis
and Meta-Analysis (Second Edition)}, Russell-Sage Foundation, New York, pp.
221-235.

\item Borenstein, M., Hedges, L., Higgins, J. \& Rothstein, H. 2010. A basic
introduction to fixed-effect and random-effects models for meta-analysis, 
\textit{Research Synthesis Methods} \textbf{1}: 97-111.

\item Branscum, A. J. \& Hanson, T. E. 2008. Bayesian nonparametric
meta-analysis using P\'{o}lya tree mixture models, \textit{Biometrics} 
\textbf{64}: 825-833.

\item Brooks, S., Gelman, A., Jones, G. \& Meng, X. 2011. \textit{Handbook
of Markov Chain Monte Carlo}, Chapman and Hall/CRC, Boca Raton, FL.

\item Burr, D. \& Doss, H. 2005. A Bayesian semiparametric model for
random-effects meta-analysis, \textit{Journal of the American Statistical
Association} \textbf{100}: 242--251.

\item Chib, S. \& Greenberg, E. 1995. Understanding the Metropolis-Hastings
algorithm, \textit{American Statistician} \textbf{49}: 327--335.

\item Cooper, H. \& Hedges, L. 2009. Research synthesis as a scientific
process, in H. Cooper, L. Hedges \& J. Valentine (eds), \textit{The Handbook
of Research Synthesis and Meta-Analysis (Second Edition)}, Russell Sage
Foundation, New York, pp. 3-18.

\item Denison, D., Holmes, C., Mallick, B. \& Smith, A. 2002. \textit{%
Bayesian Methods for Nonlinear Classification and Regression}, John Wiley
and Sons, New York.

\item DuMouchel, W. \& Normand, S.-L. 2000. Computer-modeling and graphical
strategies for meta-analysis, in D. Stangl \& D. Berry (eds), \textit{%
Meta-analysis in medicine and health policy}, Marcel Dekker, New York, pp.
127-178.

\item Egger, M., Smith, G. D., Schneider, M. \& Minder, C. 1997. Bias in
meta-analysis detected by a simple, graphical test, \textit{British Medical
Journal} \textbf{315}: 629-634.

\item Falconer, D. \& Mackay, T. 1996.\textit{\ Introduction to quantitative
genetics}, Longman, London.

\item Fleiss, J. \& Berlin, J. 2009. Effect size for dichotomous data, in H.
Cooper, L. Hedges \& J. Valentine (eds), \textit{The Handbook of Research
Synthesis and Meta-Analysis} (Second Edition), Russell Sage Foundation, New
York, pp. 237-253.

\item Gelfand, A. \& Banerjee, S. 2010. Multivariate spatial process models,
in A. Gelfand, P. Diggle, P. Guttorp \& M. Fuentes (eds), \textit{Handbook
of Spatial Statistics}, Chapman and Hall/CRC, Boca Raton, pp. 495-515.

\item Gelfand, A. \& Ghosh, J. 1998. Model choice: A minimum posterior
predictive loss approach, \textit{Biometrika} \textbf{85}: 1-11.

\item Gelfand, A. \& Smith, A. 1990. Sampling-based approaches to
calculating marginal densities, \textit{Journal of the American Statistical
Association} \textbf{85}: 398-409.

\item Gelman, A. 2006. Prior distributions for variance parameters in
hierarchical models, \textit{Bayesian Analysis} \textbf{3}: 515-533.

\item George, E. \& McCulloch, R. 1997. Approaches for Bayesian variable
selection, \textit{Statistica Sinica} \textbf{7}: 339-373.

\item Geyer, C. 1992. Practical Markov chain Monte Carlo (with discussion), 
\textit{Statistical Science} \textbf{7}: 473-511.

\item Geyer, C. 2011. Introduction to MCMC, in S. Brooks, A. Gelman, G.
Jones \& X. Meng (eds), \textit{Handbook of Markov Chain Monte Carlo}, CRC,
Boca Raton, FL, pp. 3-48.

\item Gilks, W., Wang, C., Yvonnet, B. \& Coursaget, P. 1993. Random-effects
models for longitudinal data using Gibbs sampling, \textit{Biometrics} 
\textbf{49}: 441-453.

\item Gill, J. \& Jensen, E. 1968. Probability of obtaining negative
estimates of heritability, \textit{Biometrics} \textbf{24}: 517-526.

\item Glass, G. 1976. Primary, secondary, and meta-analysis of research, 
\textit{Educational Researcher} \textbf{5}: 3-8.

\item Gleser, L. \& Olkin, I. 2009. Stochastically dependent effect sizes,
in H. Cooper, L. Hedges \& J. Valentine (eds), \textit{The Handbook of
Research Synthesis and Meta-Analysis }(Second Edition), Russell Sage
Foundation, New York, pp. 357-376.

\item Harville, D. 1977. Maximum likelihood approaches to variance component
estimation and to related problems, \textit{Journal of the American
Statistical Association} \textbf{72}: 320-338.

\item Hedges, L. 1981. Distribution theory for Glass's estimator of effect
size and related estimators, \textit{Journal of Educational and Behavioral
Statistics} \textbf{6}: 107-128.

\item Hedges, L. \& Vevea, J. 1998. Fixed and random effects models in meta
analysis, \textit{Psychological Methods} \textbf{3}: 486-504.

\item Higgins, J. \& Thompson, S. 2004. Controlling the risk of spurious
findings from meta-regression, \textit{Statistics in Medicine }\textbf{23}:
1663-1682.

\item Higgins, J., Thompson, S. \& Spiegelhalter, D. 2009. A re-evaluation
of random-effects meta-analysis, \textit{Journal of the Royal Statistical
Society}: Series A \textbf{172}: 137-159.

\item Karabatsos, G. \& Walker, S. 2012. Adaptive-modal Bayesian
nonparametric regression,\textit{\ Electronic Journal of Statistics} \textbf{%
6}: 2038-2068.

\item Konstantopoulos, S. 2007. Introduction to meta-analysis, in J.
Osbourne (ed.), \textit{Best Practices in Quantitative Methods}, Sage,
Thousand Oaks, CA, pp. 177-194.

\item Konstantopoulos, S. 2011. Fixed effects and variance components
estimation in three-level meta-analysis, \textit{Research Synthesis Methods} 
\textbf{2}: 61-76.

\item Laud, P. \& Ibrahim, J. 1995. Predictive model selection, \textit{%
Journal of the Royal Statistical Society}, \textit{Series B} \textbf{57}:
247-262.

\item Lo, A. 1984. On a class of Bayesian nonparametric estimates, \textit{%
Annals of Statistics} \textbf{12}: 351-357.

\item Louis, T. \& Zelterman, D. 1994. Bayesian approaches to research
synthesis, in H. Cooper \& L. Hedges (eds), \textit{The Handbook of Research
Synthesis}, Russell Sage Foundation, New York, pp. 411-422.

\item McCullagh, P. 1980. Regression models for ordinal data (with
discussion), \textit{Journal of the Royal Statistical Society, Series B} 
\textbf{42}: 109-142.

\item McLachlan, G. \& Peel, D. 2000. \textit{Finite Mixture Models}, John
Wiley and Sons, New York.

\item Moffitt, T. 1993. Adolescence-limited and life-course persistent
antisocial behavior: A developmental taxonomy, \textit{Psychological Review} 
\textbf{100}: 674--701.

\item Moffitt, T. 2005. The new look of behavioral genetics in developmental
psychopathology: Gene-environment interplay in antisocial behaviors, \textit{%
Psychological Bulletin} \textbf{131}: 533-554.

\item M\"{u}ller, P. \& Quintana, F. 2004. Nonparametric Bayesian data
analysis, \textit{Statistical Science} \textbf{19}: 95-110.

\item Nam, I.-S., Mengersen, K. \& Garthwaite, P. 2003. Multivariate
meta-analysis, \textit{Statistics in Medicine} \textbf{22}: 2309-2333.

\item Pinheiro, J., Bates, D., DebRoy, S., Sarkar, D. \& R Development Core
Team 2010. \textit{Nlme: Linear and Nonlinear Mixed Effects Models. R
package version 3.1-97}.

\item R Development Core Team 2012. \textit{R: A Language and Environment
for Statistical Computing}, \textit{R Foundation for Statistical Computing},
Vienna, Austria.

\item Raudenbush, S. \& Bryk, A. 2002. \textit{Hierarchical Linear Models:
Applications and Data Analysis Methods (Second Edition)}, Sage, Thousand
Oaks, CA.

\item Stevens, J. \& Taylor, A. 2009. Hierarchical dependence in
meta-analysis, \textit{Journal of Educational and Behavioral Statistics} 
\textbf{34}: 46-73.

\item Sutton, A. 2000. \textit{Methods for meta-analysis in medical research}%
, John Wiley, Chichester.

\item Talbott, B., Karabatsos, G. \& Zurheide, J. 2012. \textit{Sensitivity
of raters and the assessment of heritability of antisocial behavior: A
meta-analysis of twin, adoption, and sibling studies}, Technical report,
University of Illinois-Chicago.

\item Thompson, S. \& Higgins, J. 2002. How should meta-regression analyses
be undertaken and interpreted?, \textit{Statistics in Medicine} \textbf{21}:
1559-1573.

\item Thompson, S. \& Sharp, S. 1999. Explaining heterogeneity in
meta-analysis: A comparison of methods, \textit{Statistics in Medicine} 
\textbf{18}: 2693-2708.

\item Walker, H., Ramsey, E. \& Gresham, F. 2004. \textit{Antisocial
behavior in schools: Evidence-based practices}, Wadsworth Publishing,
Belmont, CA.
\end{description}

\bigskip

\noindent \textbf{Appendix. Behavioral-Genetic Studies}\medskip

The following numerical list provides citations for all 29
behavioral-genetic studies, which were subject to a meta-analysis. The list
is given in the order of publication year. Each item of the list presents
the identification number assigned to the given study. The identification
numbers of the studies are also shown in Figure 3.

\begin{enumerate}
\item Scarr, S. 1966. Genetic factors in activity motivation. \textit{Child
Development} \textbf{37}: 663-673.

\item Owen, D.R., \& Sines, J.O. 1970. Heritability of personality in
children. \textit{Behavior Genetics} \textbf{1}: 235-248.

\item Loehlin, J.C., \& Nichols, R.C. 1976. \textit{Heredity, environment,
and personality}. Austin: University of Texas Press.

\item Rowe, D.C. 1983. Biometrical genetic models of self-reported
delinquent behavior: A twin study. \textit{Behavior Genetics} \textbf{13}:
473-489.

\item Graham, P., \& Stevenson, J. 1985. A twin study of genetic influences
on behavioral deviance. \textit{Journal of the American Academy of Child
Psychiatry} \textbf{24}: 33-41.

\item Lytton, H., Watts, D., \& Dunn, B.E. 1988. Stability of genetic
determination from age 2 to age 9: A longitudinal twin study. \textit{Social
Biology} \textbf{35}: 62-73.

\item Eaves, L.J., Silberg, J.L., Meyer, J M., Maes, H.H., Simonoff, E.,
Pickles, A., Rutter, M., Neale, M.C., Reynolds, C.A., Erikson, M.T., Heath,
A.C., Loeber, R., Truett, K.R., \& Hewitt, J.K. 1997. Genetics and
developmental psychopathology: 2. The main effects of genes and environment
on behavioral problems in the Virginia twin study of adolescent behavioral
development. \textit{Journal of Child Psychology and Psychiatry} \textbf{38}%
: 965-980.

\item Gjone, H., \& Stevenson, J. 1997. The association between
internalizing and externalizing behavior in childhood and early adolescence:
Genetic or environmental common influences? \textit{Journal of Abnormal
Child Psychology} \textbf{25}: 277-286.

\item Eley, T.C., Lichtenstein, P., \& Stevenson, J. 1999. Sex differences
in the etiology of aggressive and nonaggressive antisocial behavior: Results
from two twin studies. \textit{Child Development} \textbf{70}: 155-168.

\item Hudziak, J.J., Rudiger, L.P., Neale, M.C., Heath, A.C., \& Todd, R.D.
2000. A twin study of inattentive, aggressive, and anxious/depressed
behaviors. \textit{Journal of the American Academy of Child and Adolescent
Psychiatry} \textbf{39}: 469-476.

\item Arseneault, L., Moffitt, T.E., Caspi, A., Taylor, A., Rijsdijk, F.V.,
\& Jaffee, S.R. et al. 2003. Strong genetic effects on cross-situational
antisocial behavior among 5-year-old children according to mothers,
teachers, examiner-observers, and twins' self reports. \textit{Journal of
Child Psychology and Psychiatry}, \textbf{44}: 832-848.

\item Eley, T.C., Lichtenstein, P., \& Moffitt, T.E. 2003. A longitudinal
behavioral genetic analysis of the etiology of aggressive and nonaggressive
antisocial behavior. \textit{Development and Psychopathology}, \textbf{15}:
383-402.

\item Van Beijsterveldt, C.E.M., Bartels, M., Hudziak, J.J., \& Boomsma,
D.I. 2003. Causes of stability of aggression from early childhood to
adolescence: A longitudinal genetic analysis in Dutch twins. \textit{%
Behavior Genetics} \textbf{33}: 591-605.

\item Vierikko, E., Pulkkinen, L., Kaprio, J., Viken, R., \& Rose, R.J.
2003. Sex differences in genetic and environmental effects on aggression. 
\textit{Aggressive Behavior} \textbf{29}, 55-68.

\item Derks, E.M., Hudziak, J.J., Van Beijsterveldt, C.E.M., Dolan, C.V., \&
Boomsma, D.I. 2004. A study of genetic and environmental influences on
maternal and paternal CBCL syndrome scores in a large sample of 3-year-old
Dutch Twins. \textit{Behavior Genetics} \textbf{34}: 571-583.

\item Gregory, A.M., Eley, T.C., \& Plomin, R. 2004. Exploring the
association between anxiety and conduct problems in a large sample of twins
aged 2-4. \textit{Journal of Abnormal Child Psychology} \textbf{32}: 111-122.

\item Van Beijsterveldt, C.E.M., Verhulst, F.C., Molenaar, P.C.M., \&
Boomsma, D.I. 2004. The genetic basis of problem behavior in 5-year-old
Dutch twin pairs. \textit{Behavior Genetics} \textbf{34}: 229-242.

\item Blonigen, D.M., Hicks, B.M., Krueger, R.F., Patrick, C.J., \& Iacono,
W.G. 2005. Psychopathic personality traits: Heritability and genetic overlap
with internalizing and externalizing psychopathology. \textit{Psychological
Medicine} \textbf{35}, 637-648.

\item Burt, S.A., McGue, M., Krueger, R.F., \& Iacono, W.G. 2005. Sources of
covariation among the child-externalizing disorders: Informant effects and
the shared environment. \textit{Psychological Medicine} \textbf{35},
1133-1144.

\item Haberstick, B.C., Schmitz, S., Young, S.E., \& Hewitt, J.K. 2005.
Contributions of genes and environments to stability and change in
externalizing and internalizing problems during elementary and middle
school. \textit{Behavior Genetics} \textbf{35}: 381-396.

\item Ligthart, L., Bartels, M., Hoekstra, R.A., Hudziak, J.J., \& Boomsma,
D.I. 2005. Genetic contributions to subtypes of aggression. \textit{Twin
Research and Human Genetics} \textbf{8}: 483-491.

\item Saudino, K.J., Ronald, A., \& Plomin, R. 2005. The etiology of
behavior problems in 7-year-old twins: Substantial genetic influence and
negligible shared environmental influence for parent ratings and ratings by
same and different teachers. \textit{Journal of Abnormal Child Psychology}, 
\textbf{33}: 113-130.

\item Tuvblad, C., Grann, M., \& Lichtenstein, P. 2006. Heritability for
adolescent antisocial behavior differs with socioeconomic status:
gene-environment interaction. \textit{Journal of Child Psychology and
Psychiatry} \textbf{47}: 734-743.

\item Vierikko, E., Pulkkinen, L., Kaprio, J., \& Rose, R.J. 2006. Genetic
and environmental sources of continuity and change in teacher-rated
aggression during early adolescence. \textit{Aggressive Behavior} \textbf{32}%
: 308-320.

\item Van Hulle, C.A., Lemery-Chalfant, K., \& Goldsmith, H.H. 2007. Genetic
and environmental influences on socio-emotional behavior in toddlers: an
initial twin study of the infant-toddler social and emotional assessment. 
\textit{Journal of Child Psychology and Psychiatry} \textbf{48}: 1014-1024.

\item Ball, H.A., Arseneault, L., Taylor, A., Maughan, B., Caspi, A., \&
Moffitt, T.E. 2008. Genetic\ and environmental influences on victims,
bullies, and bully-victims in childhood. \textit{Journal of Child Psychology
and Psychiatry}, \textbf{49}: 104-112.

\item Button, T.M. M., Lau, J.Y.F., Maughan, B., \& Eley, T.C. 2008.
Parental punitive discipline, negative life events and gene-environment
interplay in the development of externalizing behavior. \textit{%
Psychological Medicine} \textbf{38}: 29-39.

\item Saudino, K.J., Carter, A.S., Purper-Ouakil, D., \& Gorwood, P. 2008.
The etiology of behavioral problems and competencies in very young twins. 
\textit{Journal of Abnormal Psychology} \textbf{117:} 48-62.

\item Tuvblad, C., Raine, A., Zheng, M., \& Baker, L.A. 2009. Genetic and
environmental stability differs in reactive and proactive aggression. 
\textit{Aggressive Behavior} \textbf{35}: 437-452.
\end{enumerate}

\pagebreak

\begin{center}
\textbf{Figure Captions}
\end{center}

\underline{Figure 1}. Kernel density estimate of the effect size
(heritability).

\bigskip

\underline{Figure 2}. Effect size density $f(y|\mathbf{x})=f(y|\mathbf{x};%
\widehat{\sigma }=1;\boldsymbol{\zeta })$ for $\sigma _{\omega }=1/20,$ $%
1/2, $ $1,$ $2$, given $\mathbf{x}^{\intercal }\boldsymbol{\beta }=.7$ and
given sampled values of $(\mu _{j},\sigma _{j}^{2})$.

\bigskip

\underline{Figure 3}. Left panel:\ Heritability estimate, and its variance
(+), for each of the 71 independent samples of MZ and DZ twin comparisons,
provided by 29 studies. A circle refers to females, square refers to males.
Also, red refers to mother rater, black to teacher rater, blue to
self-rater, magenta to observer rater, and green to mixed raters. The study
identification number is located in the given square or circle, and this
numbering is according to the order of publication date. Right panel:\
Bayesian estimate of the heritability distribution, over the universe of
studies. Med: median; Var: variance; Skew: skewness; Kurt: kurtosis.

\bigskip

\underline{Figure 4}. For each of the 6 panels, the estimated posterior
predictive median (solid line) and interquartile range (dashed lines) of
antisocial behavior heritability (effect size), given values of a covariate,
while controlling for all other covariates by fixing their values to zero.
Mom: mother rater; Dad: father rater; Teacher: teacher rater; Self: self
rater; Observer: observer rater.

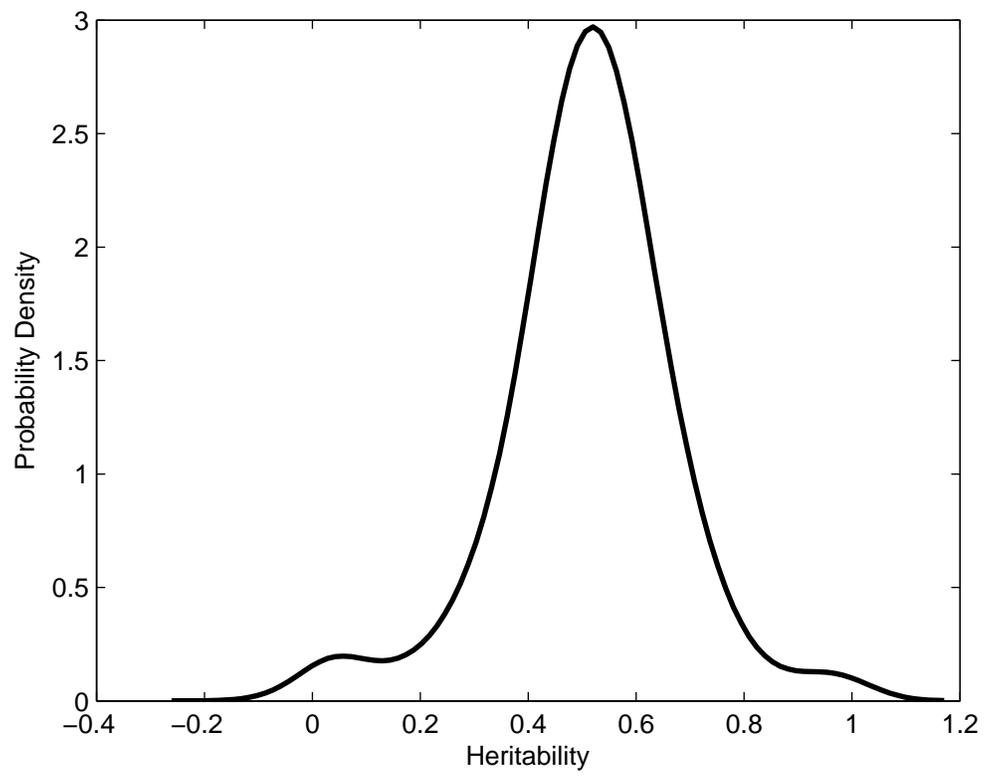
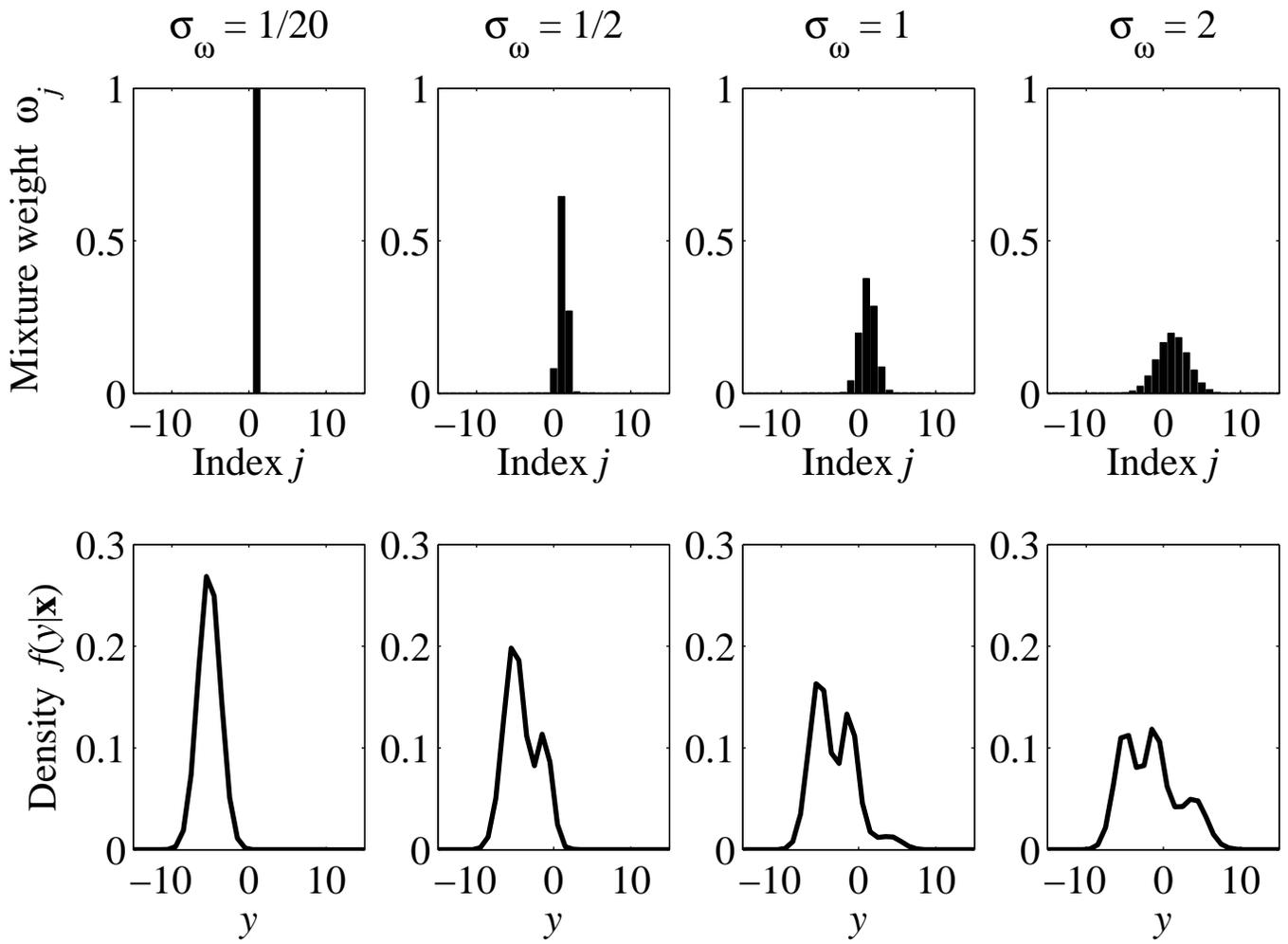
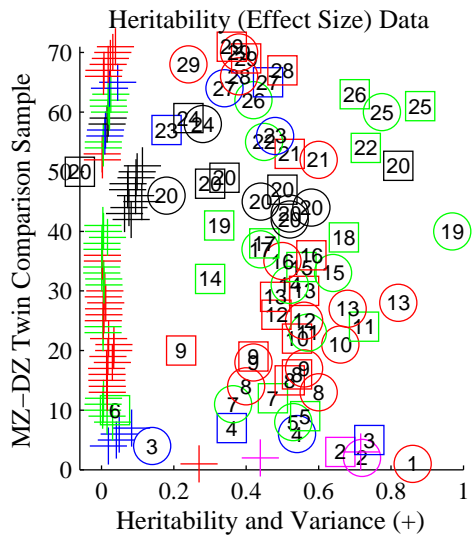
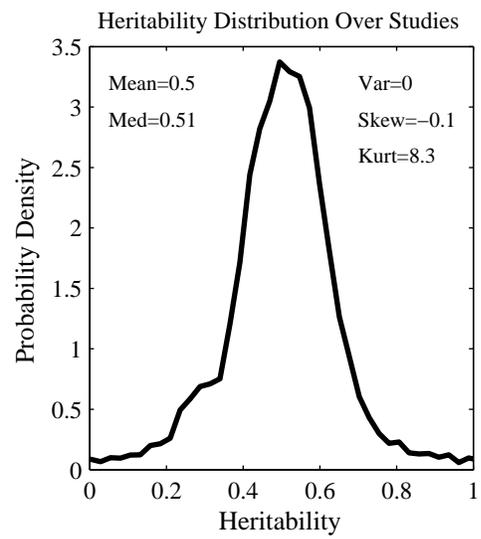
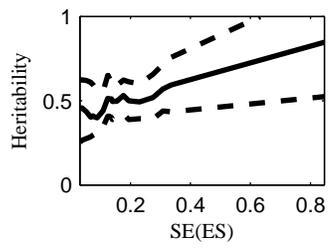
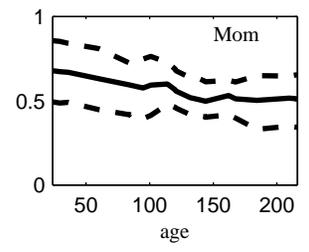
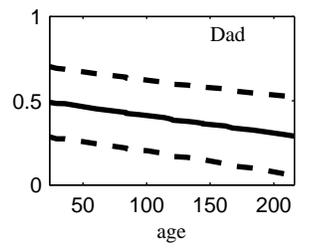
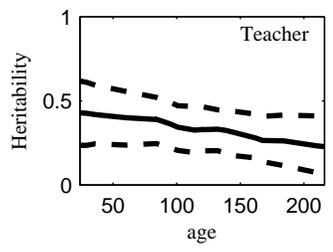
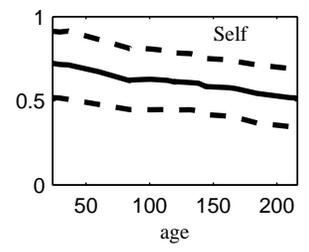
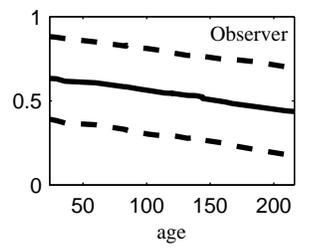
\end{document}